\documentclass[a4paper,fleqn,usenatbib]{mnras}
\usepackage{graphicx}
\usepackage{txfonts}
\usepackage{morefloats}
\usepackage{float}

\title[Abundance maps of CALIFA galaxies]
      {Oxygen abundance maps of CALIFA galaxies}

\author[I.~A.~Zinchenko et al.]
       {I.~A.~Zinchenko$^{1,2}$,  
        L.~S.~Pilyugin$^{1,2,3}$,
        E.~K.~Grebel$^{2}$, 
        S.~F.~S\'{a}nchez$^{4}$, 
        J.~M.~ V\'{\i}lchez$^{5}$   \\ 
       $^{1}$ Main Astronomical Observatory
             of National Academy of Sciences of Ukraine,
             27 Zabolotnogo str., 03680 Kiev, Ukraine \\
       $^{2}$ Astronomisches Rechen-Institut, Zentrum f\"{u}r Astronomie 
             der Universit\"{a}t Heidelberg, 
             M\"{o}nchhofstr.\ 12--14, 69120 Heidelberg, Germany \\
       $^{3}$ Kazan Federal University, 18 Kremlyovskaya St., 420008, Kazan. Russian Federation \\
       $^{4}$ Instituto de astronom\'{i}a, Universidad Auton\'{o}ma de M\'{e}xico, A.P. 70-264, 04510 M\'{e}xico, D.F. Mexico \\
       $^{5}$ Instituto de Astrof\'{\i}sica de Andaluc\'{\i}a,  CSIC, Apdo, 3004, 18080 Granada, Spain \\
}

\date{Accepted 2015 Month  00. Received 2016 January 07; in original form 2015 November 12}

\begin{document}

\maketitle

\begin{abstract}
We construct maps of the oxygen abundance distribution across the disks of 88 galaxies using CALIFA data release 2 (DR2) spectra. 
The position of the center of a galaxy (coordinates on the plate) were also taken from the CALIFA DR2. The galaxy inclination, 
the position angle of the major axis, and the optical radius were determined from the analysis of the surface brightnesses in the SDSS 
$g$ and $r$ bands of the photometric maps of SDSS data release 9. We explore the global azimuthal abundance asymmetry in the disks 
of the CALIFA galaxies and the presence of a break in the radial oxygen abundance distribution. We found that there is no significant 
global azimuthal asymmetry for our sample of galaxies, i.e., the asymmetry is small, usually lower than 0.05 dex.
The scatter in oxygen abundances around the abundance gradient has a comparable value, $\lesssim 0.05$ dex. A significant 
(possibly dominant) fraction of the asymmetry can be attributed to the uncertainties in the geometrical parameters of these galaxies.   
There is evidence for a flattening of the radial abundance gradient in the central part of 18 galaxies.
We also estimated the geometric parameters (coordinates of the center, the galaxy inclination and the position angle of the major 
axis) of our galaxies from the analysis of the abundance map. The photometry-map-based and the abundance-map-based geometrical parameters are relatively 
close to each other for the majority of the galaxies but the discrepancy is large for a few galaxies with a flat radial abundance gradient.  
\end{abstract}


\section{Introduction}

It has been well known for a long time that the disks of spiral galaxies show negative radial abundance gradients, in the sense that 
the abundance is higher at the centre and decreases with galactocentric distance \citep{Searle1971,Smith1975}.
It is a common practice to describe the nebular abundance distribution across the disk of a galaxy by the relation 
between oxygen abundance O/H and galactocentric distance $R{_g}$ and to specify this distribution by the 
characteristic abundance (the abundance at a given galactocentric distance, e.g., abundance at the centre of the disk) 
and by the radial abundance gradient. Relations of this type were determined for many galaxies by different authors 
\citep[][among many others]{VilaCostas1992,Zaritsky1994,Pilyugin2006,Pilyugin2007,Pilyugin2014a,Pilyugin2015,Moustakas2010,Gusev2012,Sanchez2014,SanchezMenguiano2016}.  
Such relations are based on the assumption that the abundance distribution across the disk is axisymmetric. 

The azimuthal abundance variations across the disk of a galaxy were discussed in a number of papers. 
The dispersion in abundance at fixed radius (the scatter around the general O/H -- $R_{g}$ trend, residuals) 
is used as the measure of the azimuthal abundance variations 
\citep[e.g.][]{Kennicutt1996,Martin1996,Bresolin2011,Li2013,Berg2015,Croxall2015}.
The number of data points in such investigations is usually small. 
The two-dimensional abundance distribution was analyzed for the galaxy NGC 628 \citep{RosalesOrtega2011}.

The observations obtained by the CALIFA survey \citep[Calar Alto Legacy Integral Field Area survey;][]{Sanchez2012} provide the possibility to construct abundance maps for disk galaxies. 
This allows one to investigate the distribution of abundances across the disk in detail, in particular the global 
azimuthal asymmetry of abundance in the disks of galaxies. We define the global azimuthal asymmetry in abundance 
in the following way. We divide the target galaxy into two semicircles and determine the difference between 
the arithmetic means of the residuals for the opposite semicircles. The differences are determined for the different 
position angles of the dividing line. The maximum absolute 
value of this difference is adopted as a measure of the global azimuthal asymmetry of abundance in the disk of 
a galaxy.  In the present work, we will determine the values of the global azimuthal abundance asymmetry in the disks of 
the CALIFA galaxies and compare them with the level of azimuthal abundance variations. 

Thus, the goal of this investigation is to construct maps of the oxygen abundance distribution in the disks 
of the selected CALIFA galaxies and to use those maps to explore the presence (or lack) of a global azimuthal 
asymmetry in the oxygen abundance as well the existence of a break in the radial oxygen abundance distribution, 
e.g., a flattening of the radial abundance gradient at the central part of a galaxy. 
Evidence for a decrease of the oxygen abundances in the central parts of 
a number of the CALIFA galaxies was found by \citet{Sanchez2014}.  

We also examine whether  the geometric parameters of a galaxy (coordinates of the center, inclination and position 
angle of the major axis) can be estimated from the analysis of the abundance map.  
The geometric parameters of a galaxy are usually determined from the analysis of the photometric 
or/and velocity map of the galaxy under the assumption that the surface brightness of the galaxy 
(or the velocity field) is a function of the radius only, i.e., that there is no azimuthal asymmetry 
in brightness (or velocity).   
Since the metallicity in the disk is also a function of the galactocentric distance one may expect 
that the abundance map can also be used for the determination of the geometric parameters 
of a galaxy. We will determine the ``chemical'' (abundance-map-based) geometrical parameters 
of our galaxies and compare them to their ``photometric'' (canonical, photometry-map-based) 
geometrical parameters. 

The paper is organized in the following way. The algorithm employed in our study is described in the Section 2. 
The results and discussion are given in Section 3. Section 4 is a brief summary.

Throughout the paper, we will use the following notations for the line fluxes: 
$R_2$ =  $I_{\rm \rm [OII] \lambda 3727+ \lambda 3729} /I_{\rm {\rm H}\beta }$, 
$N_2$  = $I_{\rm \rm [NII] \lambda 6548+ \lambda 6584} /I_{\rm {\rm H}\beta }$, 
$S_2$  = $I_{\rm \rm [SII] \lambda 6717 + \lambda 6731} /I_{\rm {\rm H}\beta }$,
$R_3$  = $I_{\rm {\rm [OIII]} \lambda 4959 + \lambda 5007} /I_{\rm {\rm H}\beta }$.
The units of the wavelengths are angstroms.

\section{Algorithm}

\subsection{The emission line fluxes}

\begin{figure*}
  \begin{center}
  \includegraphics[width=0.27\linewidth]{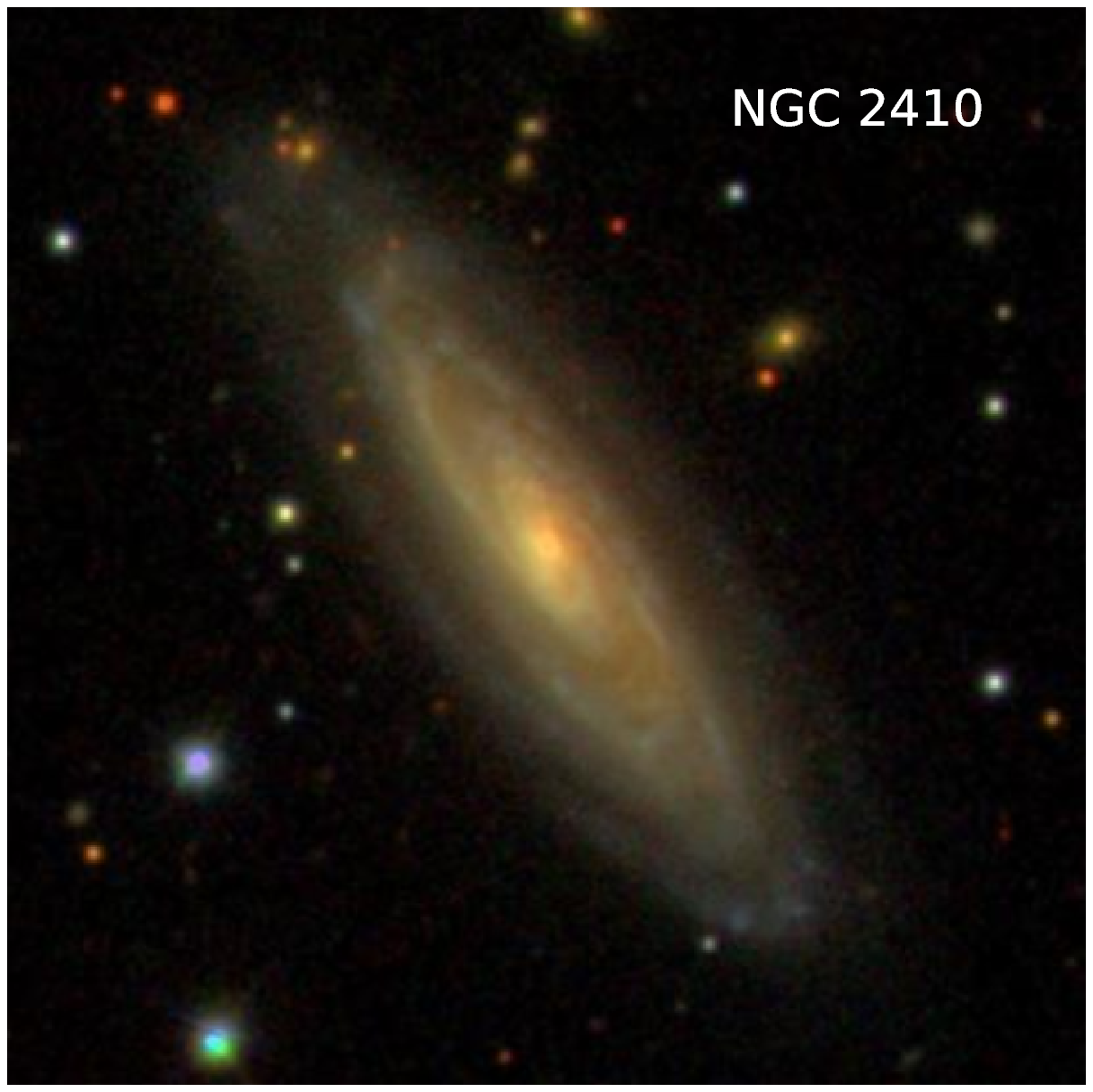}
  \includegraphics[width=0.27\linewidth]{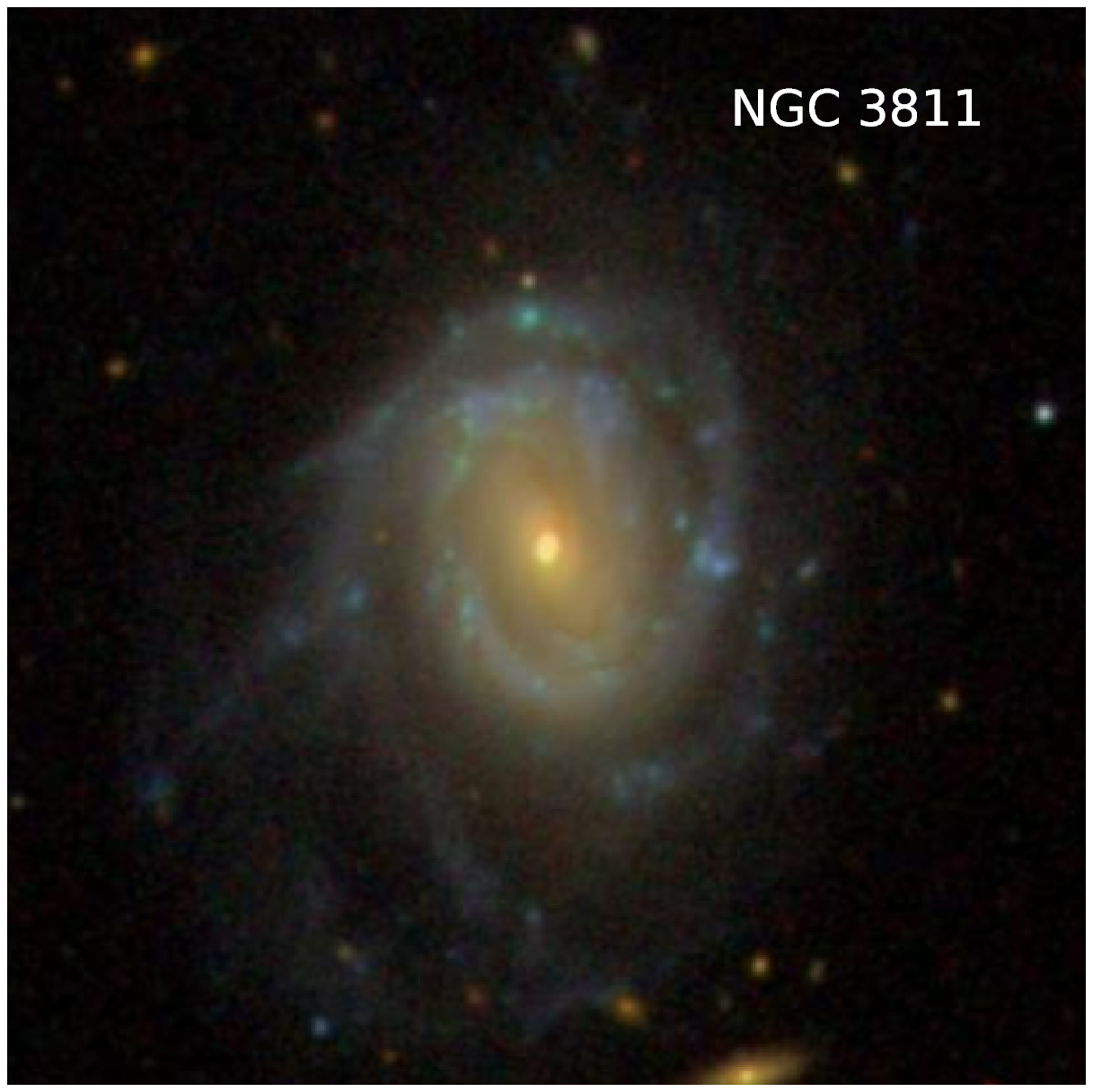}  
  \includegraphics[width=0.27\linewidth]{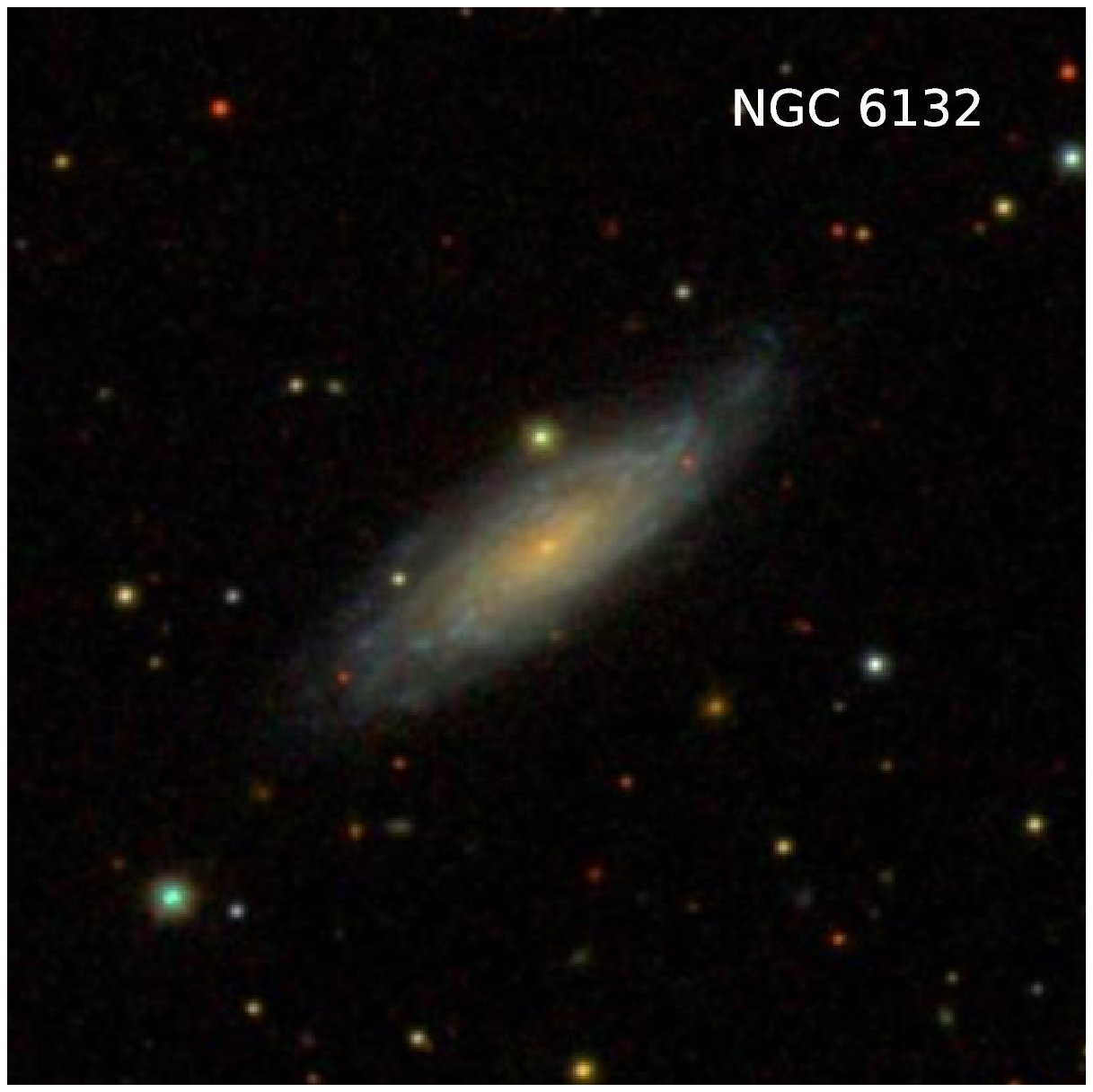}
  \includegraphics[width=0.27\linewidth]{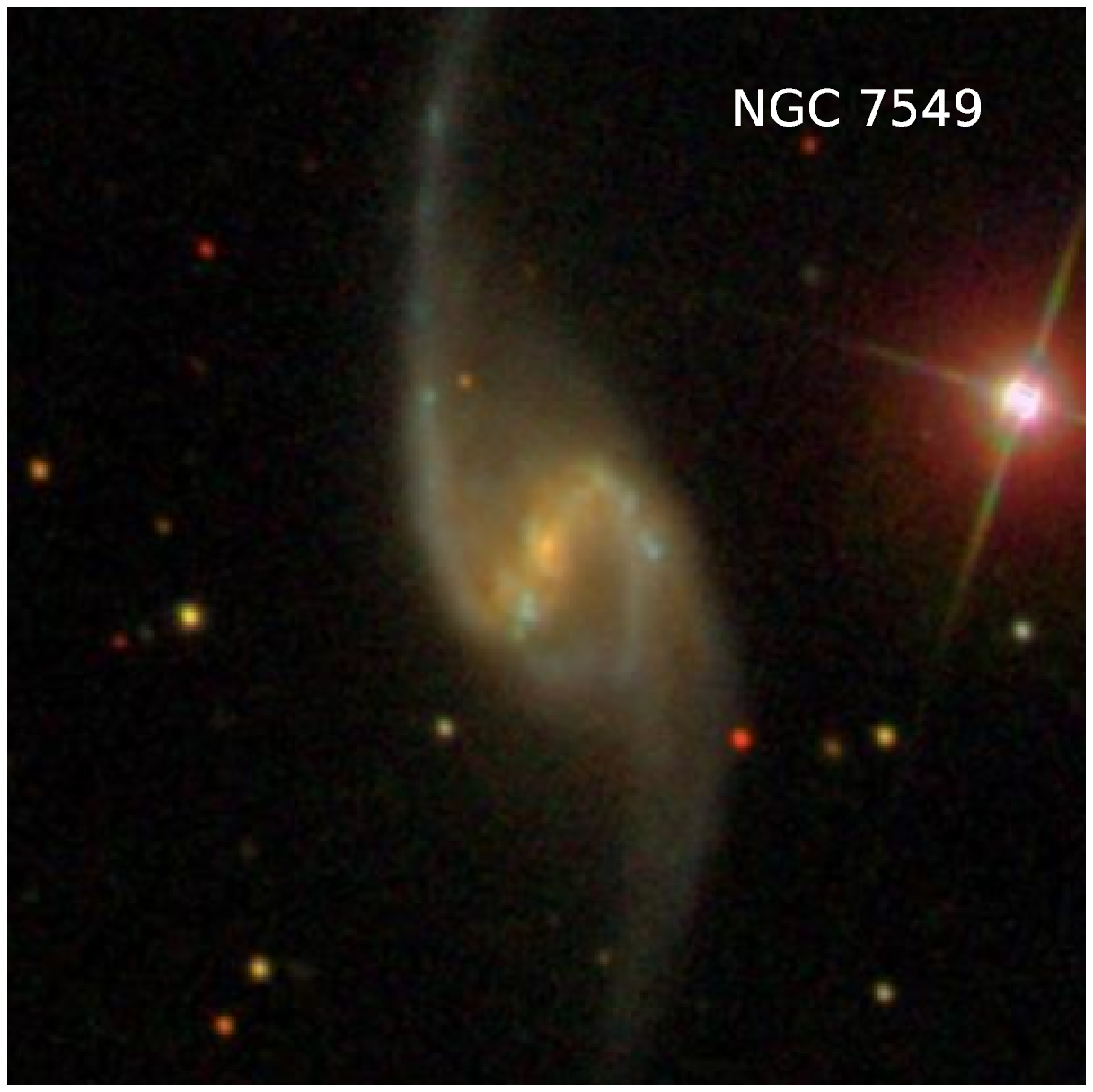}
  \includegraphics[width=0.27\linewidth]{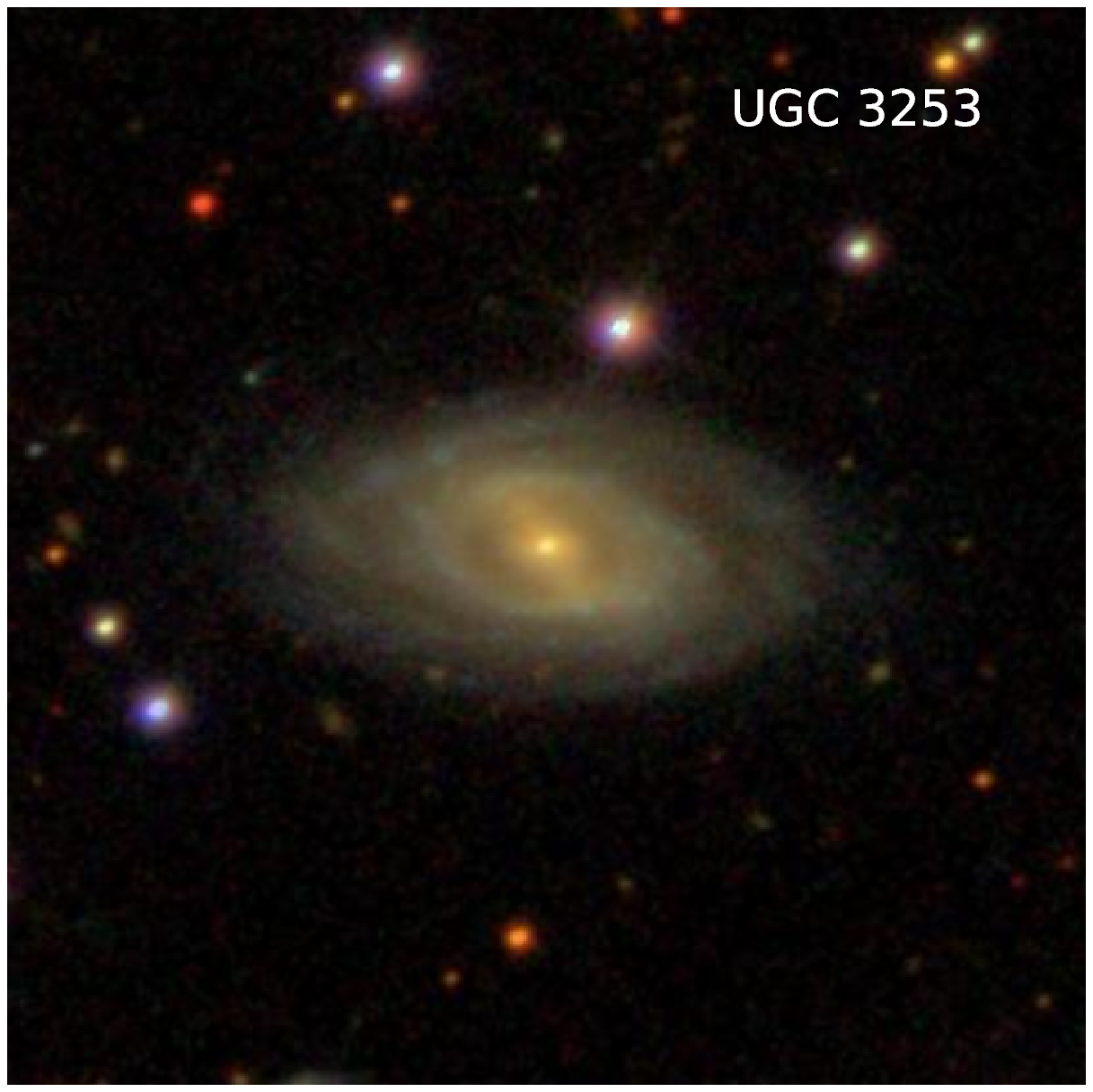}  
  \includegraphics[width=0.27\linewidth]{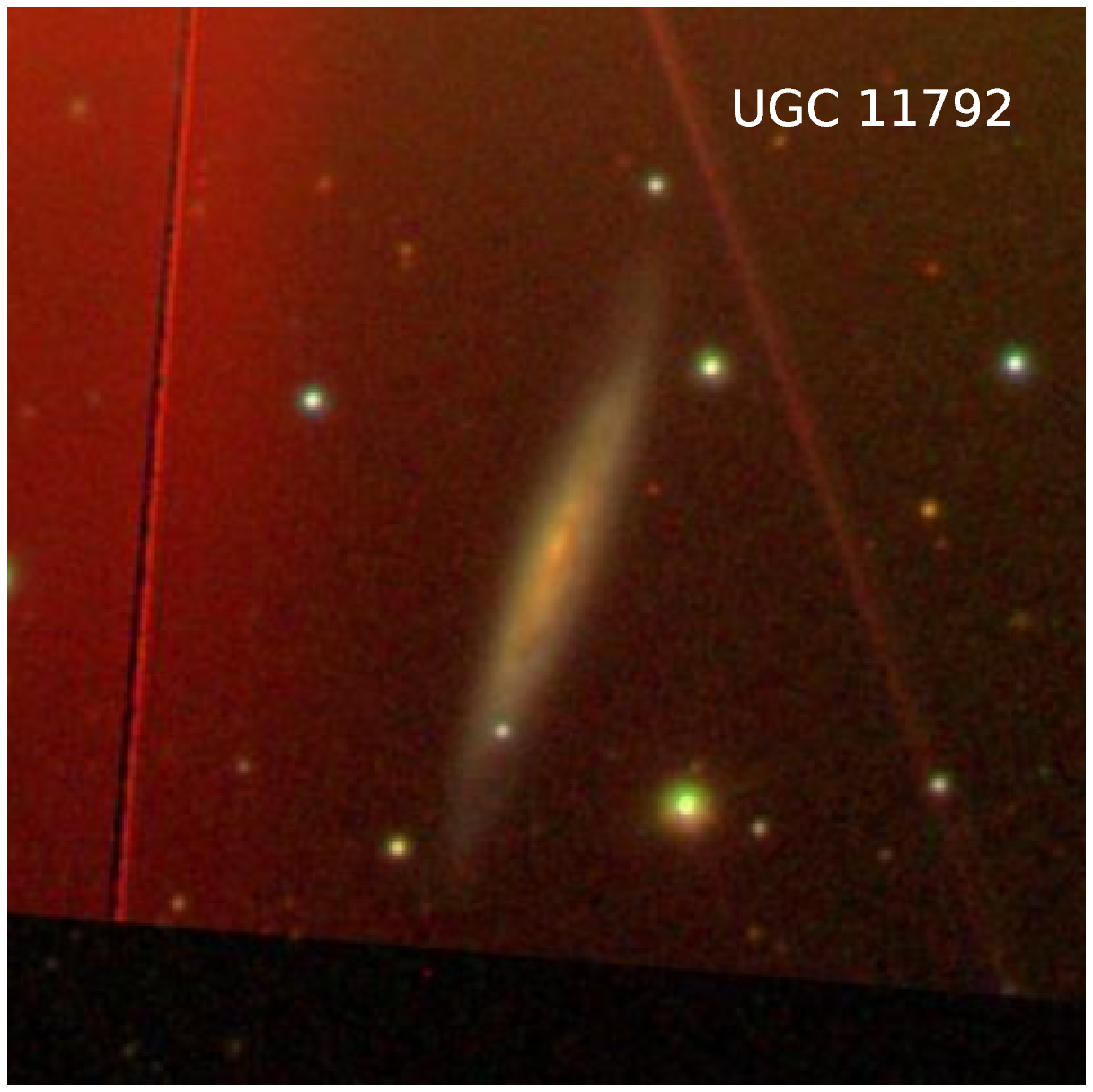}
  \end{center}
\caption{An example of {\it SDSS} images of CALIFA galaxies from our sample.
}
\label{figure:images}
\end{figure*}

We used publicly available spectra from the integral field spectroscopic CALIFA
survey data release 2 \citep[DR2;][]{Sanchez2012,Garcia-Benito2014,Walcher2014} based on
observations with the PMAS/PPAK integral field spectrophotometer mounted on the 
Calar Alto 3.5-meter telescope. CALIFA DR2 provides wide-field IFU data for 200 galaxies. 
In Fig.~\ref{figure:images} we present an example of {\it SDSS} images of six galaxies. 
The data for each galaxy consist of two spectral datacubes, which cover the spectral regions of
4300--7000~\AA\ at a spectral resolution of $R \sim 850$ (setup V500) and of 3700--5000~\AA\ 
at $R \sim 1650$ (setup V1200).

The spectrum of each spaxel from the CALIFA DR2 datacubes (setup V500) is processed into two steps. 

First, the stellar background in all spaxels is fitted using the public version of the STARLIGHT code 
\citep{CidFernandes2005,Mateus2006,Asari2007}.
We use a set of 45 synthetic simple stellar population (SSP) spectra with metallicities $Z = 0.004$, 0.02, and 0.05, 
and 15 ages from 1~Myr up to 13~Gyr  
from the evolutionary synthesis models of \citet{BC03} and the reddening law of \citet[]{CCM} with $R_V = 3.1$.
The resulting stellar radiation contribution is subtracted from the measured spectrum 
in order to find the nebular emission spectrum.
Second, the emission lines are fitted by Gaussians. The widths of the individual lines of the doublets 
[OIII]$\lambda$$\lambda$4959,5007, 
[NII]$\lambda$$\lambda$6548,6584, and 
[SII]$\lambda$$\lambda$6717,6731 are set to be equal to each other for each doublet. Since 
only the low resolution (V500 setup) spectra are used the [OII]$\lambda$$\lambda$3727,3729 
doublet is fitted by a single Gaussian. 
Applying $\chi^{2}$ minimization, the estimation of the flux error relies on the $\chi^{2}$ quadratic 
approximation in the neighborhood of the $\chi^{2}$ minimum based on a Hessian matrix.

For each spectrum, we measure the fluxes of the 
[O\,{\sc ii}]$\lambda$3727+$\lambda$3729, 
H$\beta$,  
[O\,{\sc iii}]$\lambda$4959, 
[O\,{\sc iii}]$\lambda$5007,
[N\,{\sc ii}]$\lambda$6548,
H$\alpha$,  
[N\,{\sc ii}]$\lambda$6584, 
[S\,{\sc ii}]$\lambda$6717, 
[S\,{\sc ii}]$\lambda$6731 lines.
The measured line fluxes are corrected for interstellar reddening using the theoretical H$\alpha$ to H$\beta$ ratio  
(i.e., the standard  value of H$\alpha$/H$\beta$ = 2.86) and the analytical approximation of the Whitford interstellar 
reddening law from \citet{Izotov1994}.  When the measured value of H$\alpha$/H$\beta$ is lower than 2.86 the 
reddening is adopted to be zero.

The [O\,{\sc iii}]$\lambda$5007 and $\lambda$4959 lines originate from transitions from the 
same energy level, so their flux ratio is determined only by the transition probability ratio, 
which is  very close to 3 \citep{Storey2000}. The stronger line [O\,{\sc iii}]$\lambda$5007 
is usually measured with higher precision than the weaker line [O\,{\sc iii}]$\lambda$4959. 
Therefore, the value of $R_3$ is estimated as $R_3  = 1.33$~[O\,{\sc iii}]$\lambda$5007 but not 
as a sum of the line fluxes.  
Similarly, the [N\,{\sc ii}]$\lambda$6584 and $\lambda$6548 lines also originate from transitions from the 
same energy level and the transition probability ratio for those lines is again
close to 3 \citep{Storey2000}. The value of $N_2$ is therefore estimated as 
$N_2 = 1.33$~[N\,{\sc ii}]$\lambda$6584. Thus, the lines 
H$\beta$,  
[O\,{\sc iii}]$\lambda$5007,
H$\alpha$,  
[N\,{\sc ii}]$\lambda$6584, 
[S\,{\sc ii}]$\lambda$6717, and
[S\,{\sc ii}]$\lambda$6731 
are used for the dereddening and the abundance determinations. 
The precision of the line flux is 
specified by the ratio of the flux to the flux error (parameter $\epsilon$). 
We select spectra where the parameter $\epsilon \geq 3$ for each of those lines.

\subsection{Abundance determinations}

\begin{figure}
\resizebox{0.90\hsize}{!}{\includegraphics[angle=000]{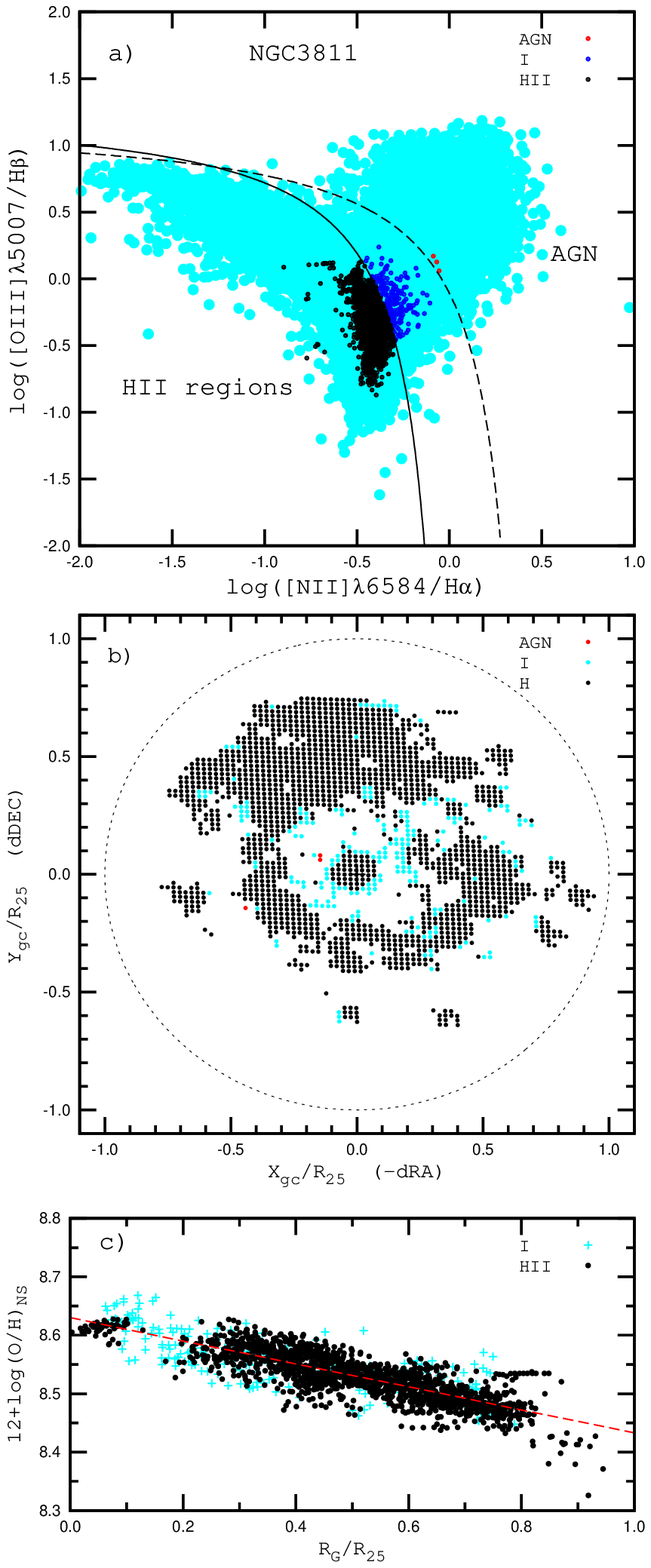}}
\caption{NGC 3811. 
Panel {\bf a} shows the BPT diagnostic diagram \citep{Baldwin1981}.  
The solid and dashed curves are the boundary between AGNs and H\,{\sc ii} regions 
defined by \citet{Kauffmann2003} and \citet{Kewley2001}, respectively.
The light-blue filled circles show the large sample of emission-line SDSS galaxies studied by \citet{Thuan2010}. 
The red points are AGN-like objects in NGC3811 according to the dividing line of \citet{Kewley2001}. 
The black points are H\,{\sc ii}-region-like objects in NGC3811 according to the dividing line of \citet{Kauffmann2003}. 
The dark-blue points are ``intermediate'' objects in NGC3811 located between the dividing lines of \citet{Kauffmann2003} and \citet{Kewley2001}.  
Panel {\bf b} shows the locations of the AGN-like, H\,{\sc ii}-region-like, and ``intermediate'' objects in NGC3811 
in the deprojected image of the galaxy. 
Panel {\bf c} shows the radial distribution of oxygen abundances in the disk of the galaxy NGC3811 
traced by the  H\,{\sc ii}-region-like (black points) and ``intermediate'' (cyan-colored crosses) objects. 
}
\label{figure:f3811}
\end{figure}

\begin{figure}
\resizebox{0.90\hsize}{!}{\includegraphics[angle=000]{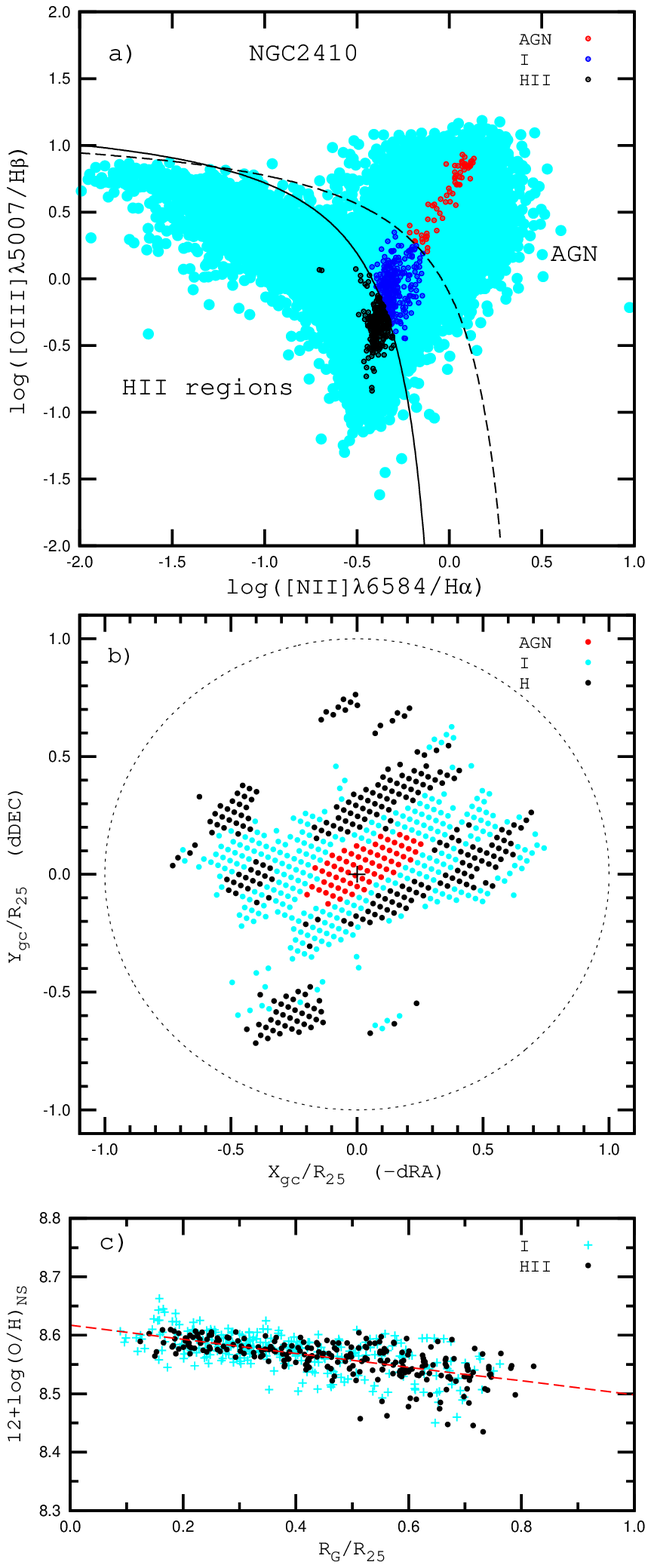}}
\caption{NGC 2410. 
The panels and symbols are the same as in Fig.~\ref{figure:f3811}.
}
\label{figure:f2410}
\end{figure}

If weak auroral lines such as such as [O\,{\sc iii}]$\lambda$4363 are detected in the spectrum of an H\,{\sc ii} region the oxygen abundance can be derived using the direct $T_e$ method, 
which is considered to yield the most reliable nebular oxygen abundance determinations \citep[e.g.,][]{Dinerstein1990}. 
This method requires the measurement of reddening-corrected diagnostic line ratios and knowledge of the local physical conditions (i.e., the effective 
tempeature of the gas, $T_e$, and its electron density).
$T_e$ is then usually determined from the intensity ratio of the auroral to nebular lines of doubly ionized oxygen;
in other words, {\em directly} from the observed line ratios (hence the name ``direct $T_e$ method'').  A short description of the required steps is given in
\citet{Dinerstein1990}.  More details can be found in \citet{Aller1984} and
\citet{Osterbrock1989}.

In our current study, the oxygen abundances will be determined through a strong-line method (the $C$ method \citep{Pilyugin2012,Pilyugin2013})  
since the auroral lines are undetectable in the majority of the CALIFA spectra \citep{Marino2013} and, consequently, the $T_e$ method cannot be applied. 
The $C$ method is based on the assumption that H\,{\sc ii} regions with a similar combination of intensities of metallicity-sensitive strong emission 
lines have similar abundances. When the strong lines [O\,{\sc iii}]$\lambda$4959,5007, [N\,{\sc ii}]$\lambda$6548,6584, and [S\,{\sc ii}]$\lambda$6717,6731
are measured in the spectrum of an H\,{\sc ii} region the oxygen (O/H)$_{C_{\rm NS}}$ abundance can be determined. When the strong lines 
[O\,{\sc ii}]$\lambda$3727,3729, [O\,{\sc iii}]$\lambda$4959,5007, and [N\,{\sc ii}]$\lambda$6548,6584 are measured in the spectrum then  we can determine 
the oxygen (O/H)$_{C_{\rm ON}}$ abundance. 

Since the oxygen line [O\,{\sc ii}]$\lambda$3727+$\lambda$3729 is not available or noisy in many spectra of the CALIFA-V500 setup the $C_{NS}$ variant of the 
$C$ method is used for the abundance determination in the H\,{\sc ii} regions. It should be emphasized that our analysis is restricted to the V500 setup data.

In order to improve the accuracy of the $C$ method we have updated our sample of 
reference H\,{\sc ii} regions that are used for the abundance determination with this method. 
New spectral measurements of 
H\,{\sc ii} regions with detected auroral lines were added to our collection \citep{Berg2012,Izotov2012,Zurita2012,Berg2013,Haurberg2013,
Li2013,Skillman2013,Brown2014,Esteban2014,Nicholls2014,Annibali2015,Berg2015,Croxall2015,Haurberg2015}. 
The abundances in those H\,{\sc ii} regions were determined through the $T_{e}$ method. 
The realization of the $T_{e}$ method, i.e., the equations of the $T_{e}$ method that serve to convert the values of the line fluxes 
to electron temperatures and abundances, are described in numerous papers \citep[e.g.,][]{Pagel1992,Izotov2006,Pilyugin2010}. The equations adopted in our current study were derived from the 
five-level atom solution with recent atomic data (Einstein coefficients for spontaneous transitions, energy level data, 
and effective cross sections for electron impact) in  \citet{Pilyugin2010}.
Using the collected data with $T_e$-based abundances we select a sample of reference H\,{\sc ii} regions following \citet{Pilyugin2013},
i.e., we select a sample of reference H\,{\sc ii} regions for which the absolute differences of the oxygen abundances 
(O/H)$_{C_{\rm ON}}$  -- (O/H)$_{T_{e}}$ and (O/H)$_{C_{\rm NS}}$  -- (O/H)$_{T_{e}}$ and of the nitrogen abundances 
(N/H)$_{C_{\rm ON}}$  -- (N/H)$_{T_{e}}$ and (N/H)$_{C_{\rm NS}}$  -- (N/H)$_{T_{e}}$ are less than 0.1 dex. 
This sample of reference H\,{\sc ii} regions contains 313  objects and will be used for the abundance determinations in the present paper. 
This sample of H\,{\sc ii} regions has been also used as calibration data in the construction of a new calibration 
for abundance determinations in  H\,{\sc ii} regions \citep{Pilyugin2016}. The good agreement between the calibration-based and $T_{e}$-based abundances 
suggests that our reference sample does indeed contain H\,{\sc ii} regions with reliable oxygen abundances (O/H)$_{T_{e}}$. 
We determine the oxygen abundance in the target object by comparing it with 30 counterpart reference H\,{\sc ii} regions in order to decrease the influence of 
the uncertainties in the oxygen abundances of the individual reference H\,{\sc ii} regions on the resulting abundance of the target objects. 

Since the abundances produced by the $C$ method and $T_{e}$ method are close to each other 
our approach is very similar to using the empirical metallicity scale defined by H\,{\sc ii} regions with oxygen 
abundances derived through the direct method ($T_e$ method).
This conclusion has been confirmed in our recent studies \citep{Pilyugin2015,Zinchenko2015}. 
It has been well known for a long time that the oxygen abundances produced by the calibrations 
based on H\,{\sc ii} regions with $T_{e}$-based measurements and calibrations based on photoionization 
models can differ by up to 0.7 dex
\citep[e.g.,][]{Pilyugin2001,Pilyugin2003b,Kewley2008,Moustakas2010,LopezSanchez2010,Bresolin2009,Bresolin2012}.  

There are different types of objects with emission-line spectra, e.g., H\,{\sc ii} region-like objects, 
AGN-like objects, and LINER-like objects. 
The intensities of the strong lines can be used to separate different types of emission-line objects according to their main  
excitation mechanism. \citet{Baldwin1981}  proposed a diagnostic diagram (the so-called BPT classification diagram) where 
the excitation properties of H\,{\sc ii} regions are established from the [N\,{\sc ii}]$\lambda$6584/H$\alpha$ 
and [O\,{\sc iii}]$\lambda$5007/H$\beta$ line ratios. 
The location of the different classes of the emission-line objects in  the [N\,{\sc ii}]$\lambda$6584/H$\alpha$ 
and [O\,{\sc iii}]$\lambda$5007/H$\beta$ and other diagrams were investigated in many studies 
\citep[e.g.,][]{Kewley2001,Kauffmann2003,Stasinska2006,Singh2013,Vogt2014,Belfiore2015,Sanchez2015b}.  
The exact location of the dividing line between H\,{\sc ii} regions and AGNs is still under debate. 

The demarcation line of \citet{Kauffmann2003} in the [N\,{\sc ii}]$\lambda$6584/H$\alpha$ vs.\ 
[O\,{\sc iii}]$\lambda$5007/H$\beta$ diagram seems to be the most stringent condition to select true  H\,{\sc ii} regions
\citep{Vogt2014}. However, an appreciable fraction of H\,{\sc ii} regions is located to the right  (thus on the ``wrong'' side)
of this demarcation line \citep{Sanchez2015b}. The goal of the current study is to construct 
metallicity maps of galaxies. Thus we are interested in using as many points as possible. 
Some  H\,{\sc ii} regions are lost when the demarcation line of \citet{Kauffmann2003} is adopted. 
Can one avoid the loss of data points with reliable $C$-based abundances?

Panel $a$ of Fig.~\ref{figure:f3811} shows the [N\,{\sc ii}]$\lambda$6584/H$\alpha$ vs.\ 
[O\,{\sc iii}]$\lambda$5007/H$\beta$ diagram for the spectra of individual spaxels in the galaxy NGC 3811.
Solid and dashed curves mark the boundary between AGNs and H\,{\sc ii} regions 
defined by \citet{Kauffmann2003} and \citet{Kewley2001}, respectively.
The light-blue filled circles show the large sample of emission-line SDSS galaxies studied by \citet{Thuan2010}. 
The red points indicate AGN-like objects in NGC3811 according to the dividing line of \citet{Kewley2001}. 
The black points represent H\,{\sc ii}-region-like objects in NGC3811 according to the dividing line of \citet{Kauffmann2003}. 
The dark-blue points are ``intermediate'' objects in NGC 3811 located between the dividing lines of \citet{Kauffmann2003} and \citet{Kewley2001}.  
Panel $b$ shows the locations of the AGN-like, H\,{\sc ii}-region-like, and ``intermediate'' objects in NGC 3811 
in the deprojected face-on image of the galaxy. 
Panel $c$ shows the radial distribution of oxygen abundances in the disk of the galaxy NGC 3811 
traced by the  H\,{\sc ii}-region-like and ``intermediate'' objects. 
Fig.~\ref{figure:f2410} shows similar diagrams for NGC 2410. The BPT diagram for NGC 2410 differs from that 
for NGC 3811. There is a well-defined AGN branch in the case of NGC 2410. 
Panels $c$ in  Fig.~\ref{figure:f3811} and Fig.~\ref{figure:f2410} clearly show that the abundances of the 
intermediate objects closely follow the general radial oxygen abundance trends traced by the  H\,{\sc ii} regions 
selected with the dividing line of \citet{Kauffmann2003}. 
Therefore the demarcation line of \citet{Kewley2001} is adopted here. 

It should be emphasized that we do not claim to establish the exact locus of the boundary between 
the H\,{\sc ii} region-like and other classes of emission line objects. We only illustrate that the $C$-based 
abundances in the objects located left of (or below) the demarcation line of \citet{Kauffmann2003} 
and the objects located between the demarcation lines of \citet{Kauffmann2003} and \citet{Kewley2001} 
show the same radial oxygen abundance trend in the disks of our target galaxies.

\subsection{The radial abundance gradient}

The galaxy inclination and position angle of the major axis are  estimated from the analysis of the surface 
brightness map in the $r$ band of the Sloan Digital Sky Survey (SDSS). We constructed radial surface brightness 
profiles in the SDSS $g$ and $r$ bands using the photometric maps of SDSS data release 9 \citep{Ahn2012} in the 
same way as in \citet{Pilyugin2014b}. The optical isophotal radius $R_{25}$ of a galaxy is determined 
from the analysis of the surface brightness profiles  in the SDSS $g$ and $r$ bands converted to the Vega $B$ band. 
We use the spaxel coordinates of the CALIFA plates. The location of the center of the galaxy is at $X_{0}$ 
and $Y_{0}$. The galactocentric $X$-coordinate is the right ascension offset with opposite sign. 
The galactocentric $Y$-coordinate is the declination offset. Since the size of a spaxel is equal to one arcsec 
the offset in spaxels is equal to the offset in arcsec. We also use fractional radii, i.e., 
radii normalized to the optical isophotal radius $R_{25}$. 

The radial oxygen abundance distribution within the optical isophotal radius of the disk, $R_{25}$, is fitted by the 
following expression, which will be referred to as the O/H -- $R_{G}$ relation: 
\begin{equation}
12+\log({\rm O/H})  = 12+\log({\rm O/H})_{0} + grad \times (R/R_{25}) ,
\label{equation:grado}
\end{equation} 
where 12 + log(O/H)$_{0}$ is the oxygen abundance at $R_{0} = 0$, i.e., the extrapolated central oxygen abundance, 
$grad$ is the slope of the oxygen abundance gradient expressed in terms of dex~$R_{\rm 25}^{-1}$, 
and $R$/$R_{\rm 25}$ is the fractional radius (the galactocentric distance normalized to the disk's isophotal 
radius $R_{25}$). 
If there is an abundance depletion in the central part of the disk then this part is 
excluded from the gradient estimation (for more details see Section~\ref{bends}).  

The radial abundance gradient obtained using the geometrical parameters of a galaxy from an analysis of 
the surface brightness map will be referred as case $A$. 
The radial abundance gradient obtained using the geometrical parameters based on an analysis of 
the abundance map will be referred as case $C$ (see below).

\subsection{The azimuthal asymmetry}

\begin{figure*}
\resizebox{0.80\hsize}{!}{\includegraphics[angle=000]{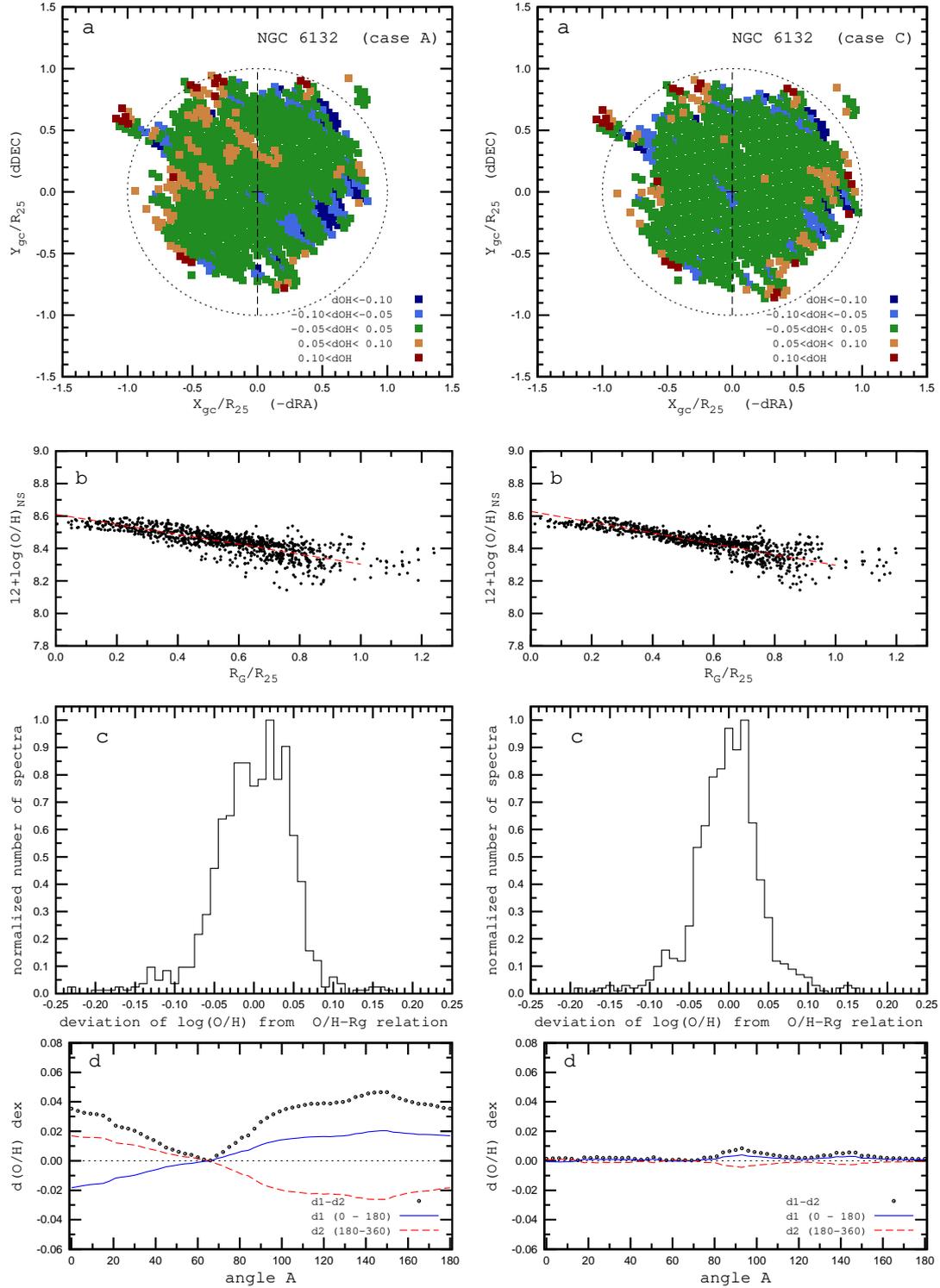}}
\caption{NGC 6132. 
The panels show the properties of the abundance distribution in the disk of the galaxy determined 
with deprojected galactocentric distances of the spaxels for the geometrical parameters of the galaxy obtained from the 
analysis of the photometric map (left column panels, case $A$) and  
with deprojected galactocentric distances of the spaxels for the geometrical parameters of the galaxy obtained from the 
analysis of the abundance  map (right column panels, case $C$). 
Panel {\bf a} in each column shows the locations of the spaxels with different abundance deviations from the O/H -- $R_g$ 
relation in the deprojected image of the galaxy. 
The dark-blue squares are spaxels with d(O/H) $\leq -0.10$ dex,
the light-blue squares stand for spaxels with $-0.10 <$ d(O/H) $\leq -0.05$ dex,
the green squares indicate spaxels with $-0.05 <$ d(O/H) $\leq 0.05$ dex,
the light-brown squares represent spaxels with $0.05 <$ d(O/H) $\leq 0.10$ dex,
and the dark-brown squares indicate spaxels with $0.10 <$ d(O/H). 
The dashed circle marks the isophotal radius $R_{25}$, and the cross marks the center of the galaxy. 
Panel {\bf b} shows the radial distribution of oxygen abundances in the disk of the target galaxy.  
The black points indicate the abundances of individual spaxels, and the red-dashed line is the linear best fit 
to those data points. 
Panel {\bf c} shows the normalized histogram of oxygen abundance deviations from the O/H -- $R_g$ relation. 
Panel {\bf d} shows the arithmetic means of the deviations from the O/H $- R_g$ relation for the spaxels within 
the sector with azimuthal angles from $A$ to $A+180$  (blue curve) and from $A+180$ to $A+360$ (red curve), 
and the absolute value of difference between those means (points) as a function of angle $A$. The dashed line 
in panel $a$ shows the dividing line for $A = 0$. 
The figures for other galaxies from our sample are available online. 
}
\label{figure:n6132ac}
\end{figure*}

We examine the global azimuthal asymmetry in the oxygen abundance distribution across the disk of a galaxy. 
We divide the galaxy into two semicircles by a dividing line at a position angle $A$.  
For a fixed value of the angle $A$, we determine 
the arithmetic mean of the deviations $d(O/H)_{1}$ from the O/H $- R_{G}$ relation for spaxels with azimuthal angles 
from $A$ to $A+180$ and the mean deviation $d(O/H)_{2}$ for spaxels with azimuthal angles from  $A+180$ to $A+360$. 
The absolute value of the difference $d(O/H) = d(O/H)_{1} - d(O/H)_{2}$ is determined for different positions of $A$. 
The position of $A$ is counted counterclockwise from the $Y$-axis (dashed line in panel $a$ of Fig.~\ref{figure:n6132ac}) 
and is changed with a step size of 3$\degr$ in the computations in both the $A$ and the $C$ cases. 
The maximum absolute value of the difference 
$d(O/H)$ is used to specify the global azimuthal asymmetry in the abundance distribution across the galaxy.
The position angle of the dividing line marking the maximum absolute value of the difference 
$d(O/H)$ will be referred to as the azimuthal asymmetry angle.

\subsection{The determination of the geometric parameters of a galaxy from the analysis 
of the abundance map}

Accurate geometric parameters of a galaxy (position of the center of a galaxy, inclination angle, 
and the position angle of the major axis) should be available for the determination of a reliable 
galactocentric distance of an H\,{\sc ii} region (spaxel). In the present study, the position of the 
center of the galaxy is described by the coordinates $X_0$ and $Y_0$ of the center of the galaxy in the CCD images.  
It is a canonical way to determine the geometric parameters of a galaxy from the analysis of the photometric 
or/and velocity map of the galaxy. It is assumed that the surface brightness distribution 
(or velocity field) is axisymmetric. 
Since the metallicity in the disk is a function of the galactocentric distance one can expect that 
the abundance map can also be used for the determination of the geometric parameters of a galaxy. 
Here we examine whether the geometric parameters of a galaxy can be estimated from the analysis 
of the abundance map.

Hence, in addition we obtain the geometric parameters of our galaxies from the analysis of the abundance map. We search for a set of four parameters:
the position of the galaxy center on the CALIFA plate (the $X_0$ and $Y_0$ coordinates in spaxels), the inclination $i$ of the galaxy, 
and the position angle $PA$ of the major axis. We determine those parameters from the requirement that the correlation coefficient 
between oxygen abundance and galactocentric distance is maximum or the scatter around O/H -- $R_{g}$ relation is minimum. It should be 
emphasized that both conditions result in the same values of the geometrical parameters. We use the data points within the optical 
radius of the galaxy. If there is a flattening of the metallicity distribution in the central part of the galaxy then this area is excluded 
from the analysis. The optical radius of a galaxy $R_{25}$ is fixed and comes from the analysis of the photometric map. 

Fig.~\ref{figure:n6132ac} illustrates the application of the described algorithm. The  galaxy NGC 6132 is a spiral 
{\sl Sab} galaxy (morphological type $T = 2.0$) at a distance of 76.5 Mpc ({\sc leda},{\sc ned}). Our analysis of 
the $r$-band SDSS map results in an inclination angle $i = 68\degr$ with a position angle of the major axis of 
$PA = 125\degr$. The angular optical radius of the galaxy is 0.56 arcmin. This results in a linear optical 
radius $R_{25} = 12.46$ kpc for the adopted distance. The left-column panels of Fig.~\ref{figure:n6132ac} show 
the properties of the abundance distribution across the disk of NGC~6132 where the deprojected radii of the H\,{\sc ii} 
regions (spaxels) were computed using the geometric parameters of the galaxy 
obtained from the photometric map, case $A$. From the analysis of the abundance map, we obtained the position of the ``chemical center'' 
of the galaxy at $X_0 = 34.3$ spaxels, $Y_0 = 34.3$ spaxels, a position angle of the major axis of $PA = 126\degr$, 
and an inclination angle $i = 69\degr$. The panels in the right column of Fig.~\ref{figure:n6132ac} show the properties 
of the abundance distribution across the disk of NGC~6132 where the deprojected radii of the H\,{\sc ii} regions 
were computed using the geometric parameters of case $C$. The abundance-map-based geometrical 
parameters will be estimated for each CALIFA galaxy of our sample below. The comparison between the photometry-map-based 
and abundance-map-based geometrical parameters can tell us something about the credibility of the abundance-map-based 
geometrical parameters of galaxies.

\section{Results}

\subsection{Our sample}
 
We consider late-type galaxies from the CALIFA DR2 list.
We include in our sample galaxies with morphological type Sa and later types 
for which the following conditions are fulfilled:
a) the number of spaxels with measured H$\beta$, [O\,{\sc iii}]$\lambda$5007, H$\alpha$,  
[N\,{\sc ii}]$\lambda$6584, [S\,{\sc ii}]$\lambda$6717, and [S\,{\sc ii}]$\lambda$6731 
emission lines is larger than a few hundreds;
b) those spaxels are distributed along the radius and with the azimuthal angle 
in such way that the radial abundance trend and azimuthal variations in the 
oxygen abundances can be investigated.
The number of the data points is usually larger than three hundred for each of the 
selected galaxies, but we also included three galaxies with 213, 299, and 176 data points, respectively. 
Of course, the procedure of including/rejecting galaxies in our sample is somewhat arbitrary.

The $C$-based abundances in some galaxies (e.g., UGC 312, UGC 7012) show unusually large scatter. 
The difference between maximum and minimum oxygen abundances at a given galactocentric distance can be as large as up to an order of 
magnitude. This  prevents a reliable determination of the radial abundance trend and the azimuthal variations in oxygen abundances. 
When we consider only the spaxels where the flux to the flux error $\epsilon > 5$ for each line (instead of 
$\epsilon > 3$) then the scatter significantly decreases. This suggests that the large scatter in the oxygen abundances 
in those galaxies seems to be artificial and can be attributed to the uncertainty in the flux measurements 
or/and to the uncertainty in the oxygen abundance determinations through the $C$ method. Those galaxies (eight galaxies) 
were excluded from further consideration. 
We also excluded three galaxies (NGC 825, NGC 2906, and UGC 3995) with very flat ($\ga -0.05$ dex)  radial abundance gradients 
where we cannot estimate the geometrical parameters of the galaxy from the abundance map. 
Our final sample contains 88 CALIFA  galaxies. 

Table \ref{table:sample} lists the general characteristics of each galaxy: morphological type, 
isophotal radius $R_{25}$, and distance.
Table \ref{table:property} lists the obtained properties of the oxygen abundance distributions 
in the disks of our sample of CALIFA galaxies for both the photometric and abundance-map derived 
geometric parameters.

\begin{figure}
\resizebox{1.00\hsize}{!}{\includegraphics[angle=000]{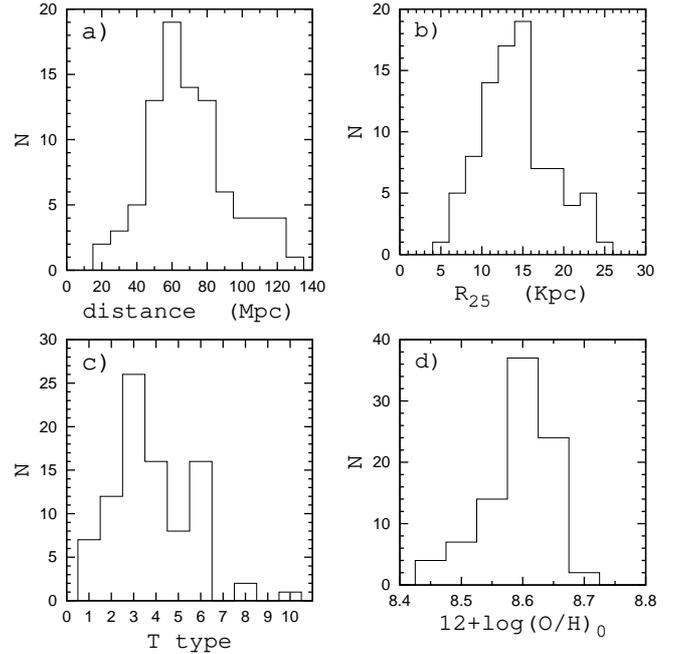}}
\caption{
Histograms of distances (panel $a$), 
optical radii $R_{25}$ (panel $b$), 
morphological $T$ types (panel $c$), and  
central oxygen abundances 12 + log(O/H)$_{0}$ (panel $d$) 
for our sample of galaxies.
}
\label{figure:gistprop}
\end{figure}

Figure~\ref{figure:gistprop} summarizes the properties of our sample of CALIFA galaxies. 
Figure~\ref{figure:gistprop} shows histograms of the distances (panel $a$), 
optical radii $R_{25}$ (panel $b$), morphological $T$ types (panel $c$), and  
central oxygen abundances 12 + log(O/H)$_{0}$ (panel $d$) for our galaxy sample.
The optical radii of our galaxies range from $\sim 5$ kpc to $\sim 25$ kpc, but galaxies with radii between 
10 and 16 kpc occur most frequently. 
The central oxygen abundances of most of the galaxies have a value between   
12 + log(O/H)$_{0} = 8.5$ and 12 + log(O/H)$_{0} = 8.7$.

\begin{figure*}
\resizebox{0.95\hsize}{!}{\includegraphics[angle=000]{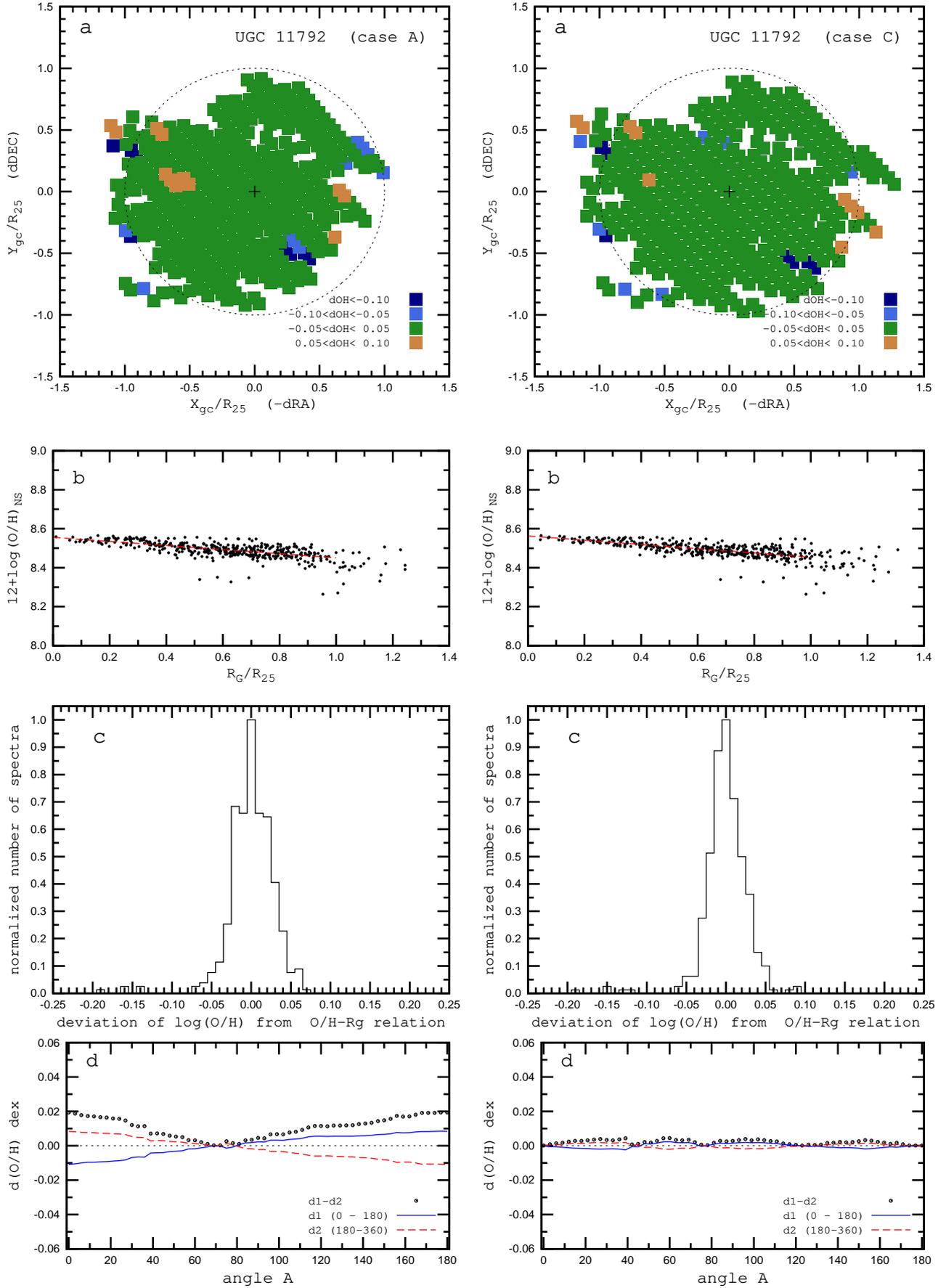}}
\caption{The galaxy UGC 11792 is a galaxy with high inclination, $i = 77\degr$. 
The panels and symbols are the same as in Fig.~\ref{figure:n6132ac}.
}
\label{figure:u11792ac}
\end{figure*}

A number of galaxies of our sample have a large inclination angle, $i \ga 70\degr$. 
Can one obtain a reliable deprojected abundance map for such galaxies?  
A satisfactory agreement between the photometry-map-based values and abundance-map-based geometrical parameters 
of such galaxies (as well as between the values of the radial abundance gradients obtained with 
photometry-map-based and abundance-map-based geometrical parameters; 
Table \ref{table:property}) can be considered as evidence in favor that reliable deprojected abundance maps 
can be obtained from the CALIFA measurements even for galaxies with an inclination up to about $80 \degr$.  
Figure~\ref{figure:u11792ac} shows the results for the galaxy UGC 11792 which is an example of a galaxy with 
high inclination, $i = 77\degr$.

\subsection{Comparison of the abundance-based and photometric-based geometrical parameters}

\begin{figure*}
\resizebox{1.00\hsize}{!}{\includegraphics[angle=000]{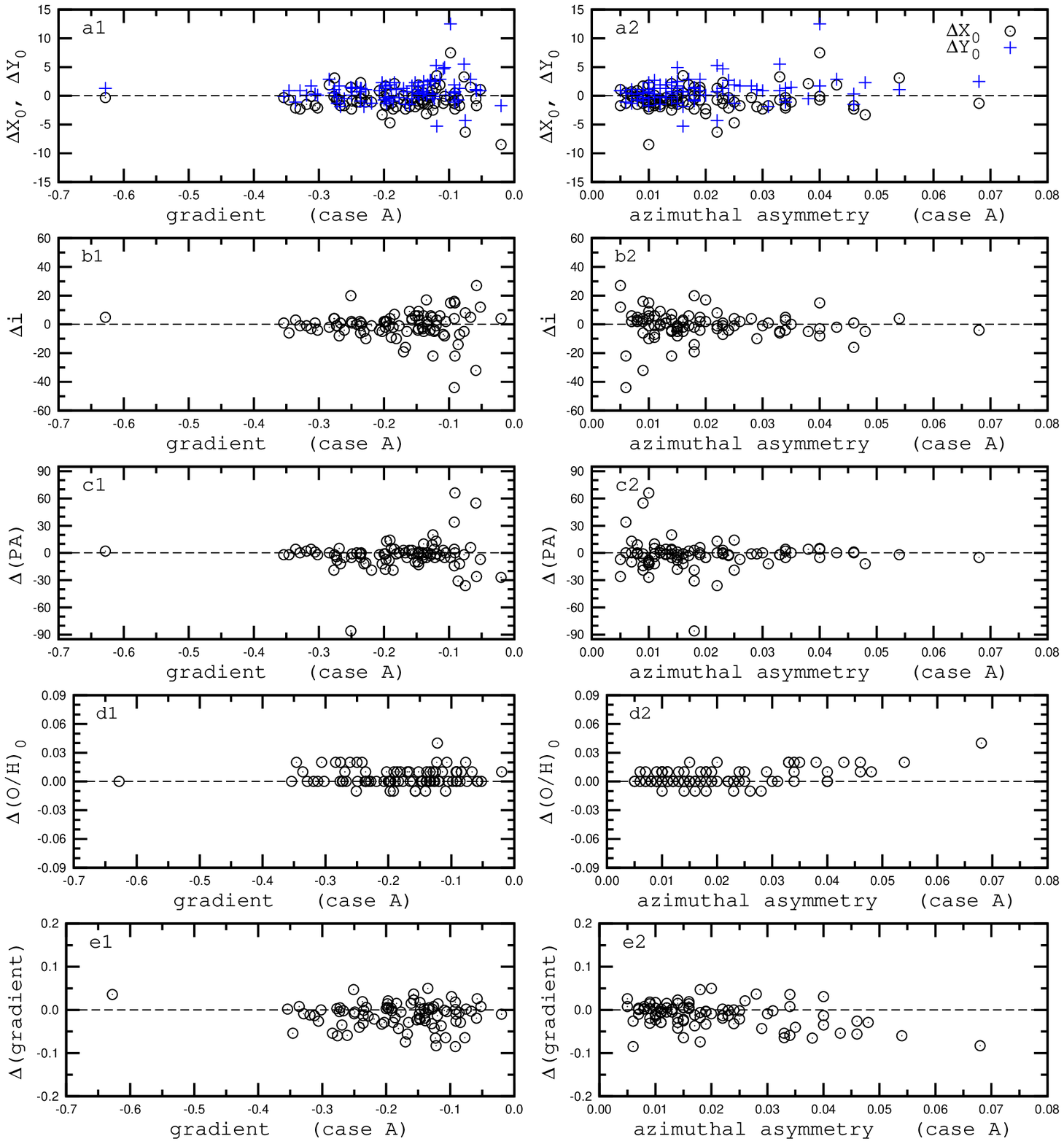}}
\caption{
The difference between the geometric parameters of a galaxy (position of the center of a galaxy
$X_{0}$, $Y_{0}$ (panels $a1,a2$), the inclination angle $i$ (panels $b1,b2$),  
and the position angle of the major axis $PA$ (panels $c1,c2$)) determined from the analysis of the abundance 
map and photometric map of the galaxy. The difference is plotted as a function of the radial oxygen abundance 
gradient across the disk (left column panels) and azimuthal asymmetry of the oxygen abundances (right column panels).
The photometry-map-based radial oxygen abundance gradient and azimuthal asymmetry value (case $A$ in the text 
and in the previous figures) are plotted along the $X$-axis. 
Panels $d1$ and $d2$ show the difference between central oxygen abundances of the disk obtained with 
abundance-map-based and photometry-map-based geometrical parameters of a galaxy as a 
function of the radial abundance gradient (panel $d1$) and azimuthal asymmetry (panel $d2$).
Panels $e1$ and $e2$ show the difference between radial oxygen abundance gradients across  the disk obtained with 
abundance-map-based and photometry-map-based geometrical parameters of a galaxy as a 
function of the radial abundance gradient (panel $e1$) and azimuthal asymmetry (panel $e2$).
}
\label{figure:grad-dx}
\end{figure*}

Panel $a1$ of Fig.~\ref{figure:grad-dx} shows the differences between the coordinates of the center 
obtained from the analysis of the photometric maps and the coordinates obtained 
from the analysis of the abundance maps  
(the differences in the $X_{0}$ coordinates are presented by circles and the difference in the $Y_{0}$ coordinates by the plus signs) 
as a function of the radial abundance gradient across the optical disk of a galaxy. 
Panel $a2$ of Fig.~\ref{figure:grad-dx} shows the differences between the coordinates of the center 
as a function of the azimuthal asymmetry in the abundance within the optical disk of a galaxy. 
The photometry-map-based radial abundance gradient and asymmetry value are plotted along the $X$-axis. 
Inspection of Fig.~\ref{figure:grad-dx} and Table \ref{table:property} show that the photometry-map-based 
and abundance-map-based coordinates of the centers of galaxies usually agree within 2 -- 3 spaxels, 
i.e., the inferred photometric and chemical centers of a galaxy are usually close to each other.
However, there is a large (more than 5 spaxels) disagreement between the photometry-map-based and 
the abundance-map-based coordinates of the centers for some galaxies with flat radial abundance gradients.

Panels $b1$ and $b2$ of Fig.~\ref{figure:grad-dx} show the comparison between the inclination angle $i_{A}$ 
obtained from the analysis of the photometric maps and the inclination angle $i_{C}$ obtained 
from the analysis of the abundance maps. 
Panels $c1$ and $c2$ of Fig.~\ref{figure:grad-dx} show the comparison between the position angle of the major axis
$PA_{A}$ obtained from the analysis of the photometric maps and the  position angle of the major 
axis $PA_{C}$ obtained from the analysis of the abundance maps. 
Inspection of Fig.~\ref{figure:grad-dx} shows that the inclination angles and the position angle of the major 
axis derived from the analysis of the abundance maps are rather close to those obtained from the 
analysis of the photometric maps for the majority of our galaxies. 
Again, there is a large (more than 10\degr) disagreement between the photometry-map-based and 
the abundance-map-based inclination angles for some galaxies with flat radial abundance gradients. 
A flat radial abundance gradient (i.e., small variations of the abundance with radius) makes the accurate 
determination of the geometrical parameters of a galaxy from the analysis of the abundance map difficult. 

The galaxy NGC 171, which has a relatively steep radial abundance gradient, $-0.25$ dex $R_{25}^{-1}$, also shows 
a large disagreement between the photometry-map-based and the abundance-map-based values of the inclination angle 
$i_{A} - i_{C}$   (panel $b1$ of Fig.~\ref{figure:grad-dx}) and between the values of the position angle of the 
major axis $PA_{A} - PA_{C}$   (panel $c1$ of Fig.~\ref{figure:grad-dx}). The inclination angle of the galaxy 
NGC 171 is small; $i_A = 20\degr$. Therefore, the determined radial abundance gradient is not very sensitive to the 
values of (and error in) the values of the inclination angle and  the position angle of the major axis. 
Hence, both the photometry-map-based and the abundance-map-based values of the inclination angle 
and of the position angle of the  major axis can be rather uncertain. 

Panels $d1$ and $d2$ show the difference between the central oxygen abundances of the disk 12+log(O/H)$_{0}$ obtained with 
photometry-map-based and abundance-map-based geometrical parameters of a galaxy as a 
function of the radial abundance gradient (panel $d1$) and asymmetry parameter (panel $d2$).
The difference in the central oxygen abundance is usually small, less than 0.02 dex.
Panels $e1$ and $e2$ show the difference between radial abundance gradients across  the disk obtained with 
photometry-map-based and abundance-map-based geometrical parameters of a galaxy as a 
function of the radial abundance gradient (panel $e1$) and asymmetry parameter (panel $e2$).
The difference is small (less than 0.05 dex $R_{25}^{-1}$) but for some galaxies it can be up to 
0.1 dex $R_{25}^{-1}$. 

Thus, the geometrical parameters of a galaxy can be estimated from the analysis of the abundance map.

\subsection{Azimuthal asymmetry of the abundances in the disk}

\begin{figure}
\resizebox{1.00\hsize}{!}{\includegraphics[angle=000]{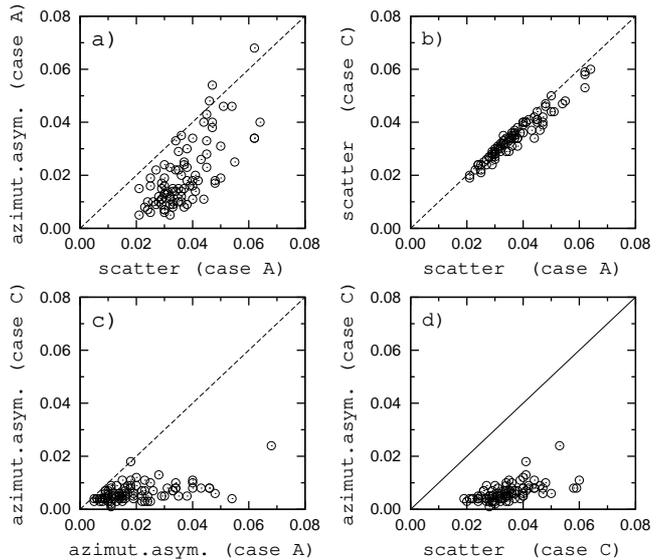}}
\caption{
Panel $a$ shows  the value (dex) of azimuthal asymmetry in the oxygen abundances in the disk of a galaxy 
as a function of the scatter (dex) in oxygen abundances around the  O/H $- R_{g}$ relation for our 
sample of galaxies for case $A$ (with geometric parameters of a galaxy determined from the analysis 
of the photometric map).
Panel $b$ shows the comparison between the values of the scatter for case $A$ and case $C$
(with geometric parameters of a galaxy determined from the analysis of the abundance map).
Panel $c$ shows the comparison between the values of the azimuthal asymmetry for case $A$ and $C$. 
Panel $d$ shows the value of azimuthal asymmetry as a function of the scatter 
for case $C$.
}
\label{figure:sigasym}
\end{figure}

Fig.~\ref{figure:n6132ac} shows the galaxy NGC 6132, which is a galaxy with possible 
azimuthal asymmetry in the oxygen abundance in the disk. 
Indeed, the upper left panel of Fig.~\ref{figure:n6132ac}  
shows that the points with positive oxygen abundance deviations (brown points) from the general radial abundance trend are 
usually located in the opposite semicircle than the points with negative oxygen abundance deviations (blue points). 
The difference between the means of the residuals in two semicircles divided by the line with an angle 
$A$ = 147$\degr$ is $\sim$0.046 dex. The scatter $\sigma$ in oxygen abundances around 
the  O/H -- $R_{g}$ relation is 0.051 dex.

Let us clarify how significant this value of azimuthal asymmetry is.
Panel $a$ of Fig.~\ref{figure:sigasym} shows the value of azimuthal asymmetry in the oxygen abundance 
in the disk of a galaxy as a function of the scatter $\sigma$ in oxygen abundances around the  O/H $- R_{g}$ relation  
for our sample of galaxies for the case $A$. The values of the global azimuthal asymmetry are small 
and exceed 0.04 dex for a few galaxies only. 

Inspection of panel $a$ of Fig.~\ref{figure:sigasym} shows that there is a correlation between the value of 
azimuthal asymmetry and the value of the azimuthal variation (scatter $\sigma$ around of the O/H $- R_{g}$ trend) 
in oxygen abundance for the case $A$. On the one hand, this correlation may suggest that the azimuthal asymmetry makes a 
significant contribution to the scatter around the O/H -- $R_{g}$ relation. This is the case if the global 
azimuthal asymmetry is real. On the other hand, the obtained azimuthal asymmetry may be artificial and 
can be caused by the uncertainties in the geometrical parameters of a galaxy or/and by the azimuthal asymmetry
of the error in oxygen abundance.  
Since the geometric parameters for case $C$ are derived by minimizing the scatter around the O/H $- R_{g}$ relation
while the geometric parameters for case $A$ are derived from the analysis of the photometric maps, the comparison 
of the azimuthal asymmetry and scatter for case $A$ and case $C$ allows us to check whether the obtained 
value of the azimuthal asymmetry for case $A$ is real.

The values of the scatter $\sigma$ in oxygen abundances around the  O/H $- R_{g}$ relation  for case $A$ 
are close to the ones for case $C$ (panel $b$ of Fig.~\ref{figure:sigasym}).
In contrast, the values of azimuthal asymmetry in the oxygen abundance for case $C$ are significantly 
lower than those for case $A$ (panel $c$ of Fig.~\ref{figure:sigasym}). 
The values of the global azimuthal asymmetry are significantly smaller than the values of the scatter $\sigma$ 
in oxygen abundances around the  O/H $- R_{g}$ relation for case $C$ (panel $d$ of Fig.~\ref{figure:sigasym}). 
Thus, the panels $b$, $c$, and $d$ of Fig.~\ref{figure:sigasym} suggest 
that the scatter $\sigma$ in oxygen abundances around the  O/H $- R_{g}$ relation 
cannot be attributed to the global azimuthal asymmetry in the oxygen abundance distribution. 
The azimuthal asymmetry obtained for case $A$ (with photometry-map-based geometrical parameters) 
seems to be artificial and may be caused by the uncertainties in the geometrical parameters of a galaxy 
at least in some galaxies. 


\begin{figure}
\resizebox{1.00\hsize}{!}{\includegraphics[angle=000]{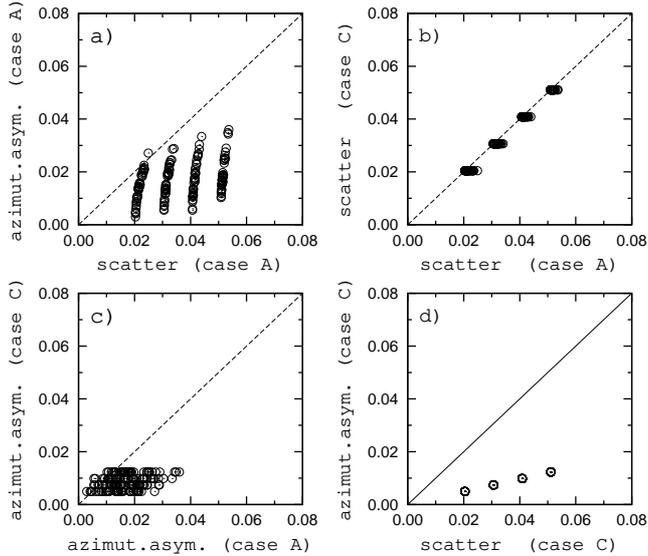}}
\caption{Simulation of the azimuthal asymmetry parameters. 
Models with abundance gradient values of $-0.15$ and 
$-0.3$~dex~$R_{\rm 25}^{-1}$ and four values of the scatter in the oxygen 
abundances, 0.02, 0.03, 0.04, and 0.05 dex, were considered.
A random deviation of the position of the galaxy center, inclination, and 
the position of the major axis was introduced to simulate the uncertainties 
in the geometrical parameters.
The panels and symbols are the same as in Fig.~\ref{figure:sigasym}.
}
\label{figure:sigasym-sim}
\end{figure}

To verify whether the values of the global azimuthal asymmetry for the case $A$ can be caused 
by uncertainties in the geometrical parameters of a galaxy obtained from the analysis of 
the photometric map, the following numerical experiment was performed. A disk model including 
400 points randomly distributed across the disk was constructed. The oxygen abundance 
in each point was determined from the adopted radial abundance gradient. Then the random 
error was added to the oxygen abundance in each point in a such way that the random 
abundance errors follow a Gaussian with an adopted scatter $\sigma$. Two values of the 
abundance gradient, $-0.15$ and $-0.3$~dex~$R_{\rm 25}^{-1}$, and four values of the scatter in oxygen 
abundances, $\sigma$ = 0.02, 0.03, 0.04, 0.05 dex, were considered. Thus, eight disk 
models were examined. 
We considered three values of the initial inclination of a galaxy, $i = 25, 45$, and $65\degr$. 
The initial position angle of the major axis of the galaxy was fixed at $PA = 45\degr$.
These models do not assume any cause of the appearance of global azimuthal asymmetry, 
except randomly distributed errors in the oxygen abundances caused by the finite number of 
points in the abundance map (discussed below). Formally we can consider this 
case as the case $C$.

We introduced a random deviation of the position of the galaxy center 
$\Delta X_0$ and $\Delta Y_0$, the galaxy inclination $\Delta i$, and the position of 
the major axis $\Delta (PA)$. The maximal values of those deviations were chosen to be  
$\Delta X_0, \Delta Y_0 = \pm5$ spaxels, $\Delta i = \pm15\degr$, and 
$\Delta (PA) = \pm20\degr$. Those deviations cover the range of 
the discrepancy between geometrical parameters in the cases $A$ and 
$C$ for the bulk of galaxies from our sample (see Fig.~\ref{figure:grad-dx}). 
Models with non-zero deviations of the geometric parameters can be 
regarded as the case $A$. 
The azimuthal asymmetry parameters were calculated for a set of 160 models.
Fig.~\ref{figure:sigasym-sim} shows the results of our simulations. 
The panels and symbols in Fig.~\ref{figure:sigasym-sim} are the same as 
in Fig.~\ref{figure:sigasym}. 
The discontinuity of the distribution of scatter is attributed to the low number of 
the initial values of the scatter; only four initial values of scatter in the oxygen abundances were considered.
The panels $a$ and $d$ of Fig.~\ref{figure:sigasym-sim}, where the lowest asymmetries of case $A$
for a given scatter value are in close agreement with the asymmetries of the case $C$,  
confirm that minimizing the scatter in oxygen abundance and deviations of the geometric 
parameters results in minimizing the azimuthal asymmetry. 
Comparison between Fig.~\ref{figure:sigasym} and Fig.~\ref{figure:sigasym-sim} 
shows clearly that our simulation reproduces the general behavior of the scatter 
and global azimuthal asymmetry obtained for our sample of the CALIFA galaxies.  
It can be considered as evidence favouring that the values of the global 
azimuthal asymmetry for the case $A$ can be caused by the rather small uncertainties 
in the geometrical parameters of a galaxy obtained from the analysis of 
the photometric map. 
The obtained range of the global azimuthal asymmetry for case $C$ can be attributed  
to the azimuthal asymmetry in the randomly distributed errors in the oxygen abundances.

\begin{figure*}
\resizebox{1.00\hsize}{!}{\includegraphics[angle=000]{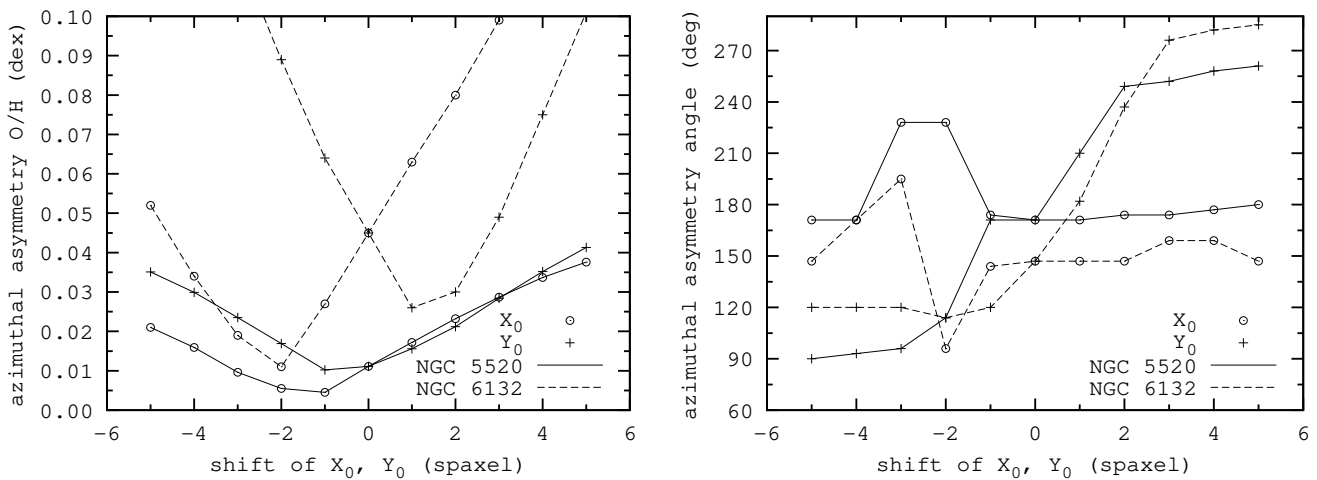}}
\caption{
The value of the global azimuthal asymmetry (left panel) and its azimuthal
angle (right panel) for the case $A$ as a function of the shift of the position of the coordinates of the centre 
($X_{O}$, $Y_{O}$) of the galaxies NGC~5520 (solid lines) and NGC~6132 (dashed lines) along the $X$ (circles) 
or $Y$ (plus signs) axes.
}
\label{figure:dxy-as}
\end{figure*}

\begin{figure}
\resizebox{1.00\hsize}{!}{\includegraphics[angle=000]{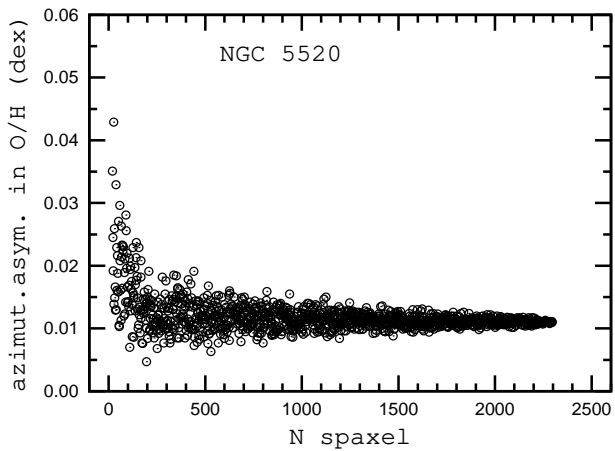}}
\caption{
The value of  the global azimuthal asymmetry as a function of the number of the spaxels in the map. 
The set of the abundance maps is constructed on the base of the map of the galaxy NGC~5520 by the 
random selection of the spaxels (see text). 
}
\label{figure:n-as}
\end{figure}

To illustrate how uncertainties in the position of the galactic center affected to the 
global azimuthal asymmetry we consider NGC~5520 and NGC~6132.
We shift the position of the galaxy's center along the $X$ or $Y$ axis (while other geometrical 
parameters are fixed) and determine the value of the global azimuthal asymmetry and its azimuthal angle.  
The left panel of Fig.~\ref {figure:dxy-as} shows the value of the global azimuthal 
asymmetry as a function of shift along the $X$ or $Y$ axis. 
This figure demonstrates that the range of the values of the global azimuthal asymmetry obtained 
for case $A$ can be reproduced by the uncertainties in the geometrical parameters of the galaxies.
The right panel of Fig.~\ref {figure:dxy-as} shows the azimuthal angle of 
the global azimuthal asymmetry as a function of the shift along the $X$ or $Y$ axis. This figure shows 
that in case of a small global azimuthal asymmetry,
small uncertainties in the position of the galactic center can lead to large uncertainties 
in the angle of the global azimuthal asymmetry.

Azimuthal variations of the oxygen abundance across galactic disks were discussed in the literature for several galaxies. 
The scatter in oxygen abundance at a given radius was usually attributed to azimuthal variations in the oxygen abundance 
distribution. 
\citet{Kennicutt1996} have considered the dispersion in abundance at fixed radius in the disk of the 
galaxy M101 using 41 H\,{\sc ii} regions. They found that there is evidence in favor of a non-axisymmetric abundance distribution. 
However, they noted that more data are needed to test whether this asymmetry is real. 
\citet{Li2013} obtained and analyzed the radial oxygen abundance gradient in the disk of M101 using 
a sample of 79 H\,{\sc ii} regions. They found no evidence for significant azimuthal variations of the oxygen 
abundance across the entire disk of this galaxy.  

\citet{RosalesOrtega2011} found scatter around the O/H -- $R_{g}$ relation of 0.128 dex in their fiber-to-fiber 
sample and of 0.070 dex in their H\,{\sc ii} region catalog of NGC 628. They concluded that the physical conditions and the star 
formation history of different symmetric regions of the galaxy NGC 628 have evolved in a different manner. 
\citet{Berg2015} detected the auroral lines and measured direct abundances in 45 H\,{\sc ii} regions in the disk 
of the galaxy NGC 628. They found that the O/H abundances have a large intrinsic dispersion of $\sim 0.165$ dex. They 
posit that this dispersion represents an upper limit to the true dispersion in oxygen abundance at a fixed 
galactocentric distance and that some of that dispersion is caused by systematic uncertainties in the temperature 
measurements. 

\citet{Rosolowsky2008} found that there is substantial scatter of 0.11 dex in the metallicity at any given radius in the disk 
of the galaxy M33. \citet{Bresolin2011} found that the oxygen abundance gradient in the inner 2 kpc of the M33 disk 
measured from the detection of the [O\,{\sc iii}]$\lambda$4363 auroral line displays a scatter of approximately 0.06 dex. 
A large sample of  H\,{\sc ii} regions 
assembled from the literature results in a comparably small scatter (0.05 -- 0.07 dex) over the whole optical disk of M33. 
\citet{Bresolin2011} noted that this dispersion can be explained simply by the measurement uncertainties. He concluded that no 
evidence is therefore found for significant azimuthal variations in the present-day metallicity of the interstellar 
medium of M33 on spatial scales from $\sim 100$ pc to a few kpc. 
  
\citet{Croxall2015} have measured direct gas-phase abundances in 29 individual H\,{\sc ii} regions in the disk 
of the galaxy NGC5194 (M51). They found an oxygen abundance gradient with very little scatter ($\sigma \la 0.08$ dex). 
They concluded that most of this scatter can be attributed to random errors and is not caused by an intrinsic dispersion. 

\citet{Sanchez2015b} selected and investigated 396 H\,{\sc ii} regions in the galaxy NGC 6754. They found evidence of an 
azimuthal variation in the oxygen abundance of about 0.05 dex, which may be related to radial migration. 

The scatter around the O/H -- $R_{g}$ relation obtained here for the CALIFA galaxies is in the range of $\sim 0.02$ to 
$\sim 0.06$ dex, which is lower than the scatter in galaxies (from $\sim 0.05$ to $\sim 0.165$ dex) determined in the above 
quoted studies. This discrepancy may be caused by the fact that we select the spectra included in the 
construction and analysis of the metallicity maps, i.e., we use only spectra where the ratio of the flux to the flux error 
is higher than three for each of the lines used in the abundance determinations. 

Can the number of data points influence the obtained value of the global azimuthal asymmetry in a galaxy? We consider the galaxy 
NGC~5520 to examine this problem. The abundance map of NGC~5520 consists of 2316 spaxels and has a global 
azimuthal asymmetry of 0.011 dex in case $A$. We construct a set of 1000 abundance maps of the galaxy NGC~5520 with a 
reduced number of spaxels, randomly selected from the full map of 2316 spaxels.  The value of the 
global azimuthal asymmetry was determined for each constructed abundance map. Fig.~\ref{figure:n-as} shows the value of the global 
azimuthal asymmetry as a function of the number of the spaxels in the map. This figure shows that a
global azimuthal asymmetry larger than 0.02 dex appears only when the number of spaxels decreases to $\sim 200$. Since the number of 
spaxels in our target galaxies are usually higher than 200 (see Table \ref{table:property}), the obtained values of the global 
azimuthal asymmetry in the target galaxies cannot be attributed to too small a number of data points in the maps. 

Thus, there is no significant global azimuthal asymmetry for our sample of the CALIFA galaxies. 
The values of the global azimuthal asymmetry are small and can be attributed to uncertainties 
in the geometrical parameters of our galaxies.

\subsection{Bends in the radial abundance gradients} \label{bends}

\begin{figure*}
\resizebox{0.945\hsize}{!}{\includegraphics[angle=000]{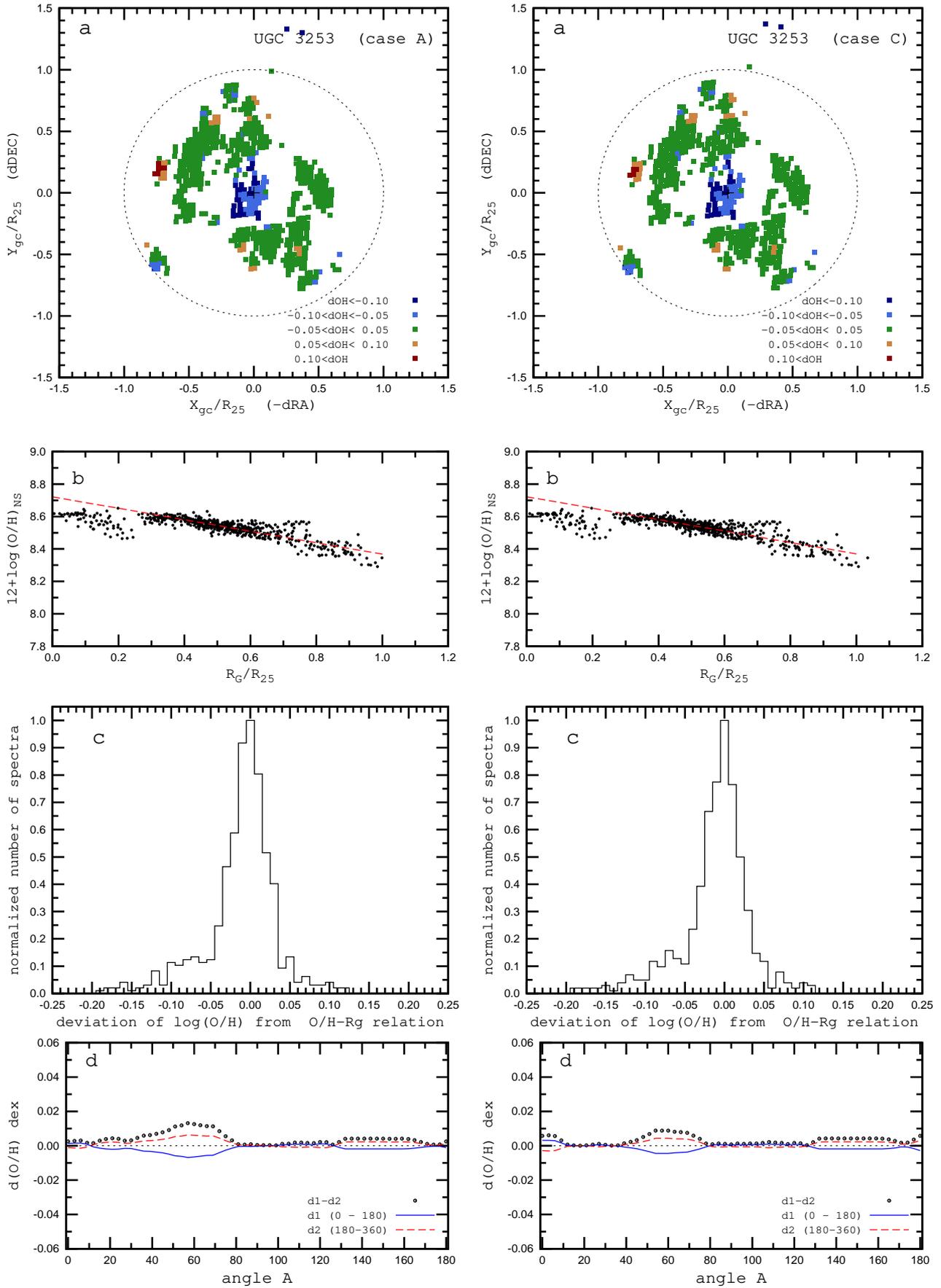}}
\caption{UGC 3253 is a galaxy with a possible bend in the 
radial abundance distribution. The oxygen abundances in the central part of the galaxy 
are systematically lower as compared to the general radial abundance trend. 
The panels and symbols are the same as in Fig.~\ref{figure:n6132ac}.
}
\label{figure:u3253ac}
\end{figure*}

We have found that the radial abundance distribution across the disks of the majority of the 
galaxies of our sample can be well fitted by a single relation. 
In a subset of galaxies a bend in the radial abundance distribution across the disk may exist within 
the optical $R_{25}$ radius.  
Fig.~\ref{figure:u3253ac} shows UGC 3253, which is a galaxy with a possible bend in the 
radial abundance distribution within the optical radius. The oxygen abundances in the central part of the galaxy 
are systematically lower as compared to the general radial abundance trend. 
Possible bends in the radial abundance distributions in other galaxies 
(a flattening or lowering in the central part) 
are reported in  Table \ref{table:property} (column 11). 
The decrease of the oxygen abundances in the central parts of 
a number of the CALIFA galaxies was first revealed by \citet{Sanchez2014}.  
The nearly flat distribution in the innermost ($\la$0.2$R_{25}$) part of the galaxy NGC 628 
was reported by \citet{RosalesOrtega2011}.

It should be noted that it is difficult to establish the exact value of the bend point due to the scatter in 
oxygen abundance values at any fixed radius and because the overall decrease of the oxygen abundance in the central 
part of a galaxy is small, within $\sim 0.1$ dex. 
Therefore, the break radius reported in  Table \ref{table:property} is an indicator that there is evidence 
for a decrease in the oxygen abundance in the central part of the galaxy rather than the exact 
value of the bend point. 

The arguments for or/and against the existence of the break in the radial abundance gradients in the 
disks of spiral galaxies were discussed in many papers 
\citep[][among others]{Vilchez1988,VilaCostas1992,Vilchez1996,Pilyugin2003a,Pilyugin2004,Bresolin2009,Bresolin2012,Goddard2011,Marino2012,Marino2016}  
There is consensus is that the change in the radial abundance gradient can occur near the isophotal 
radius of a galaxy. The depressed gas metallicity in the central part of some CALIFA galaxies noticed 
by \citet{Sanchez2014} is confirmed here. Further investigations will strengthen or reject this picture. 

Thus, there is evidence for a small change in the slope of the gradient 
(a flattening or decrease in the central part) in the disks of a number of galaxies.

\section{Summary}

We construct maps of the oxygen abundances across the disks of 88 CALIFA galaxies. The oxygen 
abundances were determined through the $C$ method using CALIFA DR2 spectra. 
Hence, here we use the empirical metallicity scale defined by H\,{\sc ii} regions 
with oxygen abundances derived through the direct method ($T_e$ method).  
The position of the center of a galaxy (coordinates on the plate) were taken 
from the CALIFA DR2. The galaxy inclination and position angle of the major axis 
were determined from our analysis of $r$ band photometric maps of SDSS data release 9. 
The optical radii were determined from the radial surface brightness profiles in the SDSS $g$ and $r$ bands 
constructed on the basis of the photometric maps and converted to the Vega $B$ band. 

We examine the global azimuthal asymmetry of the abundances in the disks of our target galaxies. 
The arithmetic mean of the deviations from the O/H $- R_{G}$ relation $d(O/H)_{1}$  for spaxels with azimuthal angles 
from $A$ to $A+180$ and mean deviation $d(O/H)_{2}$ for spaxels with azimuthal angles from  $A+180$ to $A+360$ 
are determined for different positions of $A$. 
The maximum absolute value of the  difference $d(O/H)_{1} - d(O/H)_{2}$ is used to specify the 
global azimuthal asymmetry in the abundance distribution across the galaxy.

The scatter around the O/H $- R_{g}$ relation for our sample of CALIFA galaxies is in the 
range of $\sim 0.02$ to $\sim 0.06$ dex.
There is no significant global azimuthal asymmetry for our CALIFA sample. The values of the global azimuthal 
asymmetry are small, less than 0.05 dex for the bulk of target galaxies. These values can be attributed to the uncertainties in the 
photometry-map-based  geometrical parameters of the galaxies, i.e., the uncertainties in the photometry-map-based  geometrical 
parameters of a galaxy can make an appreciable (and possibly dominant) contribution to the obtained values 
of the azimuthal asymmetry.

We have found that the radial abundance distribution across the disks of the majority of 
the galaxies of our sample can be well fitted by a single relation. 
However, in eighteen of the galaxies in our sample, the oxygen abundances in the central part of the galaxies
are systematically lower as compared to the general radial abundance trend. 
Although the decrease is rather well defined, its value is small, within $\sim$0.1 dex, and 
can be questioned taking into account the uncertainties of the abundances in the reference  H\,{\sc ii} regions. 
We note that the decrease of the oxygen abundances in the central parts of 
a number of the CALIFA galaxies was first revealed by \citet{Sanchez2014} and recently 
confirmed by \citet{SanchezMenguiano2016}.  

We estimated the geometric parameters of our galaxies (coordinates of the center, inclination and position 
angle of the major axis) from the analysis of the abundance map. The geometrical parameters are 
determined on the condition that the coefficient of the correlation between oxygen abundance and galactocentric 
distance is maximized or the scatter around the O/H -- $R_{g}$ relation is minimized. Both these conditions 
result in the same values of the geometrical parameters.  The photometry-map-based and the 
abundance-map-based geometrical parameters are relatively close to each other for the majority of our galaxies 
but the discrepancy is large for a few galaxies with a flat radial abundance gradient.

\section*{Acknowledgements}

L.S.P., E.K.G., and I.A.Z.\  acknowledge support within the framework
of Sonderforschungsbereich SFB 881 on ``The Milky Way System''
(especially subproject A5), which is funded by the German Research
Foundation (DFG). \\ 
L.S.P.\ and I.A.Z.\ thank for the hospitality of the
Astronomisches Rechen-Institut at Heidelberg University  where part of
this investigation was carried out. \\
L.S.P.\ and I.A.Z.\ acknowledge the support of the Volkswagen Foundation 
under the Trilateral Partnerships grant No. 90411. \\
I.A.Z.\ also acknowledges the special support by the NASU under the 
Main Astronomical Observatory GRID/GPU computing cluster {\tt golowood} 
project. \\
This work was partly funded by a subsidy allocated to the Kazan Federal 
University for the state assignment in the sphere of scientific 
activities (L.S.P.).  \\ 
S.F.S.\ thanks the CONACYT-125180 and DGAPA-IA100815 projects for
providing him support in this study. \\
This study uses data provided by the Calar Alto Legacy Integral Field Area 
(CALIFA) survey (http://califa.caha.es/). 
Based on observations collected at the Centro Astron\'{o}mico Hispano Alem\'{a}n (CAHA) 
at Calar Alto, operated jointly by the Max-Planck-Institut f\"{u}r Astronomie and 
the Instituto de Astrof\'{i}sica de Andaluc\'{i}a (CSIC). \\ 
This research made
use of Montage, funded by the National Aeronautics and Space
Administration's Earth Science Technology Office, Computational
Technnologies Project, under Cooperative Agreement Number NCC5-626
between NASA and the California Institute of Technology. The code is
maintained by the NASA/IPAC Infrared Science Archive. \\ 
Funding for
the SDSS and SDSS-II has been provided by the Alfred P. Sloan
Foundation, the Participating Institutions, the National Science
Foundation, the U.S. Department of Energy, the National Aeronautics
and Space Administration, the Japanese Monbukagakusho, the Max Planck
Society, and the Higher Education Funding Council for England. The
SDSS Web Site is http://www.sdss.org/.

\clearpage 

\appendix

\section{Tables}

Table \ref{table:sample} lists the general characteristics of each galaxy. \\ 
Column 1 gives its name. We have used the most widely used name for each galaxy. 
The galaxies are listed in the order of the name category, with the following
categories in descending order:  \\
NGC -- New General Catalogue,       \\
IC -- Index Catalogue,     \\
UGC -- Uppsala General Catalog of Galaxies,   \\
PGC -- Catalogue of Principal Galaxies.  \\
Columns 2 and 3  report the morphological type of the galaxy and the morphological type code $T$
from {\sc leda}.    \\
Column 4 lists the isophotal radius $R_{25}$ in arcmin of each galaxy. We determined the isophotal radius 
from the photometric maps in the SDSS $g$ and $r$ bands.   \\
Column 5 gives the isophotal radius in kpc, estimated from the data in columns 4 and 6.   \\
Column 6 reports the {\sc ned} distances using flow corrections for Virgo, the Great
Attractor, and Shapley Supercluster infall. \\

Table \ref{table:property} lists the obtained properties of the oxygen abundance distributions 
in the disks of our sample of CALIFA galaxies.  
The inferred properties of each galaxy are given in two consecutive rows. In the first row, we report 
the geometrical parameters of the galaxies obtained from the analysis of the photometric map 
and the properties of the abundance distribution in the disk of the galaxies determined 
with deprojected galactocentric distances of spaxels for those geometrical parameters (case $A$ in the text and figures). 
In the second row, we report 
the geometrical parameters of the galaxies obtained from the analysis of the abundance map 
and the properties of the abundance distribution in the disk of the galaxy determined 
with deprojected galactocentric distances of spaxels for those geometrical parameters (case $C$ in the text). 
Column 1 gives the name of the galaxy. We have used the most widely used name for each galaxy.   
The galaxies are listed in the same order as in Table \ref{table:sample}. \\
Columns 2 and 3  report the position of the galaxy center on the CALIFA exposure (the $X_{0}$ and $Y_{0}$ coordinates 
in spaxels).    \\
Columns 4 and 5 give the galaxies' inclination and the position angle of the major axes.    \\ 
Columns 6 and 7 list the extrapolated central 12 + log(O/H)$_{R_{0}}$ oxygen
abundance and the radial oxygen abundance gradient expressed in terms of dex~$R_{\rm 25}^{-1}$.  \\
Column 8 reports the scatter of oxygen abundances around the general radial oxygen
abundance trend  within the optical $R_{25}$ radius of a galaxy.   \\ 
Columns 9 and 10  give the global azimuthal asymmetry (maximum difference between the arithmetic means of the deviations from the O/H $- R_g$ 
relation for the opposite semicircle sectors) and the position of the dividing line (see panels $d$ in Fig.~\ref{figure:n6132ac}).    \\ 
Column 11 lists the break radius if a bend in the radial abundance distribution exists. \\
Column 12 provides the number of the points in the abundance map for a galaxy. 

\setcounter{table}{0}
\begin{table}
\caption[]{\label{table:sample}
The adopted general properties of our sample of the CALIFA galaxies 
}
\begin{center}
\begin{tabular}{llccrc} \hline \hline
Name                                          &
Type                                          &  
T-type                                        &
$R_{25}$                                       &
$R_{25}$                                       &
Distance                                      \\  
                                       &
                                       &
                                       &
arcmin                                 &
kpc                                    &
Mpc                                    \\  
 \hline 
NGC 1            &  Sb        &    3.1   &    0.64  &    11.47   &    61.6   \\  
NGC 36           &  SABb      &    3.0   &    0.95  &    22.38   &    81.0   \\  
NGC 171          &  Sab       &    2.2   &    0.99  &    15.21   &    52.8   \\  
NGC 180          &  Sc        &    4.6   &    1.09  &    22.39   &    70.6   \\  
NGC 192          &  SBa       &    1.1   &    0.84  &    13.51   &    55.3   \\  
NGC 237          &  SABc      &    4.5   &    0.75  &    12.20   &    55.9   \\  
NGC 444          &  Sc        &    6.4   &    0.60  &    11.33   &    64.9   \\  
NGC 477          &  Sc        &    5.0   &    0.78  &    17.95   &    79.1   \\  
NGC 496          &  Sbc       &    4.0   &    0.69  &    16.16   &    80.5   \\  
NGC 776          &  SABa      &    2.5   &    0.77  &    14.67   &    65.5   \\  
NGC 1542         &  Sab       &    2.0   &    0.60  &     8.66   &    49.6   \\  
NGC 2253         &  Sc        &    5.8   &    0.77  &    11.49   &    51.3   \\  
NGC 2347         &  Sb        &    3.1   &    0.84  &    15.39   &    63.0   \\  
NGC 2410         &  Sb        &    3.0   &    0.87  &    16.97   &    66.3   \\  
NGC 2730         &  Sd        &    8.0   &    0.78  &    12.87   &    56.7   \\  
NGC 2906         &  Sc        &    5.9   &    0.84  &     8.19   &    33.5   \\  
NGC 2916         &  Sb        &    3.1   &    1.07  &    17.43   &    56.0   \\  
NGC 3381         &  SBb       &    3.2   &    0.93  &     7.79   &    22.8   \\  
NGC 3614         &  SABc      &    5.2   &    1.41  &    15.75   &    38.4   \\  
NGC 3811         &  SBc       &    5.8   &    0.86  &    12.40   &    49.0   \\  
NGC 3994         &  Sc        &    4.9   &    0.57  &     8.19   &    49.4   \\  
NGC 4185         &  SBbc      &    3.7   &    1.06  &    18.81   &    61.0   \\  
NGC 4210         &  Sb        &    3.0   &    0.86  &    10.81   &    43.2   \\  
NGC 4470         &  Sa        &    1.4   &    0.73  &     8.15   &    38.4   \\  
NGC 4644         &  Sb        &    3.1   &    0.52  &    11.18   &    73.9   \\  
NGC 5205         &  Sbc       &    3.5   &    0.87  &     7.82   &    30.9   \\  
NGC 5406         &  Sbc       &    3.9   &    0.92  &    21.14   &    79.0   \\  
NGC 5520         &  Sb        &    3.1   &    0.80  &     7.73   &    33.2   \\  
NGC 5614         &  Sab       &    1.7   &    1.09  &    19.47   &    61.4   \\  
NGC 5720         &  Sb        &    3.0   &    0.68  &    22.37   &   113.1   \\  
NGC 5888         &  Sbc       &    3.8   &    0.60  &    22.11   &   126.7   \\  
NGC 5947         &  SBbc      &    3.6   &    0.61  &    15.62   &    88.0   \\  
NGC 6004         &  Sc        &    4.9   &    0.90  &    15.92   &    60.8   \\  
NGC 6063         &  Sc        &    5.9   &    0.75  &    10.19   &    46.7   \\  
NGC 6132         &  Sab       &    2.0   &    0.56  &    12.46   &    76.5   \\  
NGC 6154         &  Sa        &    1.0   &    0.74  &    19.09   &    88.7   \\  
NGC 6478         &  Sc        &    4.9   &    0.81  &    23.05   &    98.0   \\  
NGC 6978         &  Sb        &    2.7   &    0.69  &    17.18   &    85.6   \\  
NGC 7311         &  Sab       &    2.0   &    0.81  &    14.80   &    62.8   \\  
NGC 7321         &  SBb       &    3.1   &    0.72  &    20.50   &    97.9   \\  
NGC 7466         &  Sb        &    3.1   &    0.58  &    17.26   &   102.3   \\  
NGC 7489         &  Sc        &    6.4   &    0.85  &    21.04   &    85.1   \\  
NGC 7549         &  Sc        &    5.9   &    0.83  &    15.65   &    64.8   \\  
NGC 7591         &  SBbc      &    3.6   &    0.79  &    15.54   &    67.6   \\  
NGC 7625         &  Sa        &    1.0   &    0.81  &     5.58   &    23.7   \\  
NGC 7631         &  Sb        &    3.1   &    0.77  &    11.54   &    51.5   \\  
NGC 7653         &  Sb        &    3.1   &    0.77  &    13.06   &    58.3   \\  
NGC 7716         &  Sb        &    3.0   &    0.96  &     9.94   &    35.6   \\  
NGC 7819         &  Sb        &    3.1   &    0.78  &    15.25   &    67.2   \\  
IC  674          &  Sab       &    2.0   &    0.59  &    18.64   &   108.6   \\  
IC  1199         &  Sbc       &    3.7   &    0.66  &    14.11   &    73.5   \\  
IC  1256         &  Sb        &    3.3   &    0.69  &    14.47   &    72.1   \\  
IC  1528         &  SBb       &    3.1   &    0.95  &    14.26   &    51.6   \\  
IC  2487         &  Sb        &    3.1   &    0.78  &    14.82   &    64.3   \\  
UGC 5            &  Sbc       &    3.9   &    0.64  &    18.23   &    97.9   \\  
UGC 809          &  Sc        &    5.9   &    0.49  &     8.07   &    56.6   \\  
UGC 1057         &  Sbc       &    4.0   &    0.55  &    13.58   &    84.9   \\  
UGC 1938         &  Sbc       &    4.0   &    0.52  &    12.83   &    84.8   \\ 
\hline
\end{tabular}\\
\end{center}
\begin{flushleft}
\end{flushleft}
\end{table}

\setcounter{table}{0}
\begin{table}
\caption[]{
Continued
}
\begin{center}
\begin{tabular}{llccrc} \hline \hline
Name                                          &
Type                                          &  
T-type                                        &
$R_{25}$                                       &
$R_{25}$                                       &
Distance                                      \\  
                                       &
                                       &
                                       &
arcmin                                 &
kpc                                    &
Mpc                                    \\  
 \hline 

UGC 2403         &  SBa       &    1.3   &    0.64  &    10.20   &    54.8   \\  
UGC 3107         &  Sbc       &    4.0   &    0.45  &    14.69   &   112.2   \\  
UGC 3253         &  Sb        &    3.0   &    0.70  &    12.12   &    59.5   \\  
UGC 3969         &  Sc        &    5.9   &    0.44  &    14.45   &   112.9   \\  
UGC 4132         &  Sbc       &    4.0   &    0.73  &    15.71   &    74.0   \\  
UGC 5396         &  Sc        &    5.8   &    0.65  &    15.05   &    79.6   \\  
UGC 5598         &  Sbc       &    3.8   &    0.47  &    11.36   &    83.1   \\  
UGC 8107         &  IB        &    9.9   &    0.70  &    24.33   &   119.5   \\  
UGC 8778         &  Sab       &    2.0   &    0.51  &     7.80   &    52.6   \\  
UGC 9067         &  Sab       &    2.0   &    0.56  &    18.95   &   116.3   \\  
UGC 9476         &  SABc      &    5.2   &    0.79  &    12.02   &    52.3   \\  
UGC 9665         &  Sbc       &    4.0   &    0.65  &     8.04   &    42.5   \\  
UGC 9873         &  Sc        &    5.3   &    0.46  &    11.28   &    84.3   \\  
UGC 9892         &  Sb        &    3.0   &    0.51  &    12.63   &    85.1   \\  
UGC 10331        &  Sb        &    3.1   &    0.58  &    11.34   &    67.2   \\  
UGC 10384        &  Sab       &    1.6   &    0.46  &    10.24   &    76.5   \\  
UGC 10710        &  Sb        &    3.0   &    0.57  &    20.05   &   120.9   \\  
UGC 10811        &  Sab       &    2.0   &    0.52  &    18.83   &   124.5   \\  
UGC 10972        &  Sc        &    5.9   &    0.65  &    13.35   &    70.6   \\  
UGC 11262        &  Sc        &    6.4   &    0.51  &    12.18   &    82.1   \\  
UGC 11649        &  SBa       &    1.0   &    0.77  &    12.39   &    55.3   \\  
UGC 11792        &  Sc        &    5.8   &    0.47  &     9.27   &    67.8   \\  
UGC 11982        &  SBc       &    6.0   &    0.40  &     7.82   &    67.2   \\  
UGC 12185        &  SBab      &    2.5   &    0.67  &    17.76   &    91.1   \\  
UGC 12224        &  Sc        &    5.9   &    0.87  &    12.33   &    48.7   \\  
UGC 12519        &  SBbc      &    4.5   &    0.68  &    11.85   &    59.9   \\  
UGC 12816        &  SABc      &    5.8   &    0.65  &    13.60   &    71.9   \\  
UGC 12864        &  SBb       &    3.1   &    0.74  &    13.69   &    63.6   \\  
PGC 1841         &  SBa       &    1.0   &    0.77  &    10.62   &    47.4   \\  
PGC 64373        &  Sd        &    8.0   &    0.63  &    15.01   &    81.9   \\  

                                                             \hline
\end{tabular}\\
\end{center}
\begin{flushleft}
\end{flushleft}
\end{table}

\clearpage

\setcounter{table}{1}
\begin{table*}
\caption[]{\label{table:property}
The obtained properties of the oxygen abundance distributions in the disks of the sample of CALIFA galaxies 
with available abundance maps.
}
\begin{center}
\begin{tabular}{lllcrccccccr} \hline \hline
Name                                          &
\multicolumn{2}{c}{Center location}           &  
$i$                                          &
P.A.                                          &
(O/H)$_{0}^{a}$                                & 
O/H gradient                                  &
$\sigma$(O/H)                                 &
\multicolumn{2}{c}{Maximum azimuthal asymmetry}  &  
Break                                        &
Number of                                    \\
                                        &
$X_{0}$                                  &
$Y_{0}$                                  &
                                        &
                                        &
                                        & 
                                        &
                                        &
amplitude                               &
angle                                   &
radius                                  &
spectra                                 \\

                             &
spaxel                       &
spaxel                       &
degree                       &
degree                       &
                             &
dex $R_{25}^{-1}$              &
dex                          & 
dex                          &
degree                       &
$R_{25}$                      &
                             \\  \hline 
1                            &
2                            &
3                            &
4                            &
5                            &
6                            &
7                            &
8                            & 
9                            &
10                           &
11                           &
12                           \\  \hline   
NGC 1            &  38   &    36    &   45  &    101   &    8.61  &   -0.092  &   0.029  &   0.009  &  138  &        &    951  \\  
                 &  36.7 &    36.5  &   61  &     87   &    8.61  &   -0.074  &   0.024  &   0.004  &   12  &        &         \\  
NGC 36           &  39   &    32    &   63  &     14   &    8.62  &   -0.107  &   0.033  &   0.015  &  102  &        &    524  \\  
                 &  38.9 &    36.9  &   55  &      9   &    8.64  &   -0.171  &   0.035  &   0.011  &   75  &        &         \\  
NGC 171          &  35   &    34    &   20  &    161   &    8.67  &   -0.251  &   0.040  &   0.018  &    6  &  0.3   &    480  \\  
                 &  36.5 &    35.5  &   40  &     75   &    8.66  &   -0.204  &   0.039  &   0.011  &    6  &        &         \\  
NGC 180          &  35   &    32    &   46  &    165   &    8.64  &   -0.198  &   0.037  &   0.015  &  165  &        &    906  \\  
                 &  33.5 &    33.3  &   44  &      3   &    8.64  &   -0.192  &   0.036  &   0.006  &  120  &        &         \\  
NGC 192          &  38   &    33    &   65  &    168   &    8.64  &   -0.148  &   0.030  &   0.007  &   54  &        &    553  \\  
                 &  38.7 &    34.1  &   71  &    158   &    8.64  &   -0.158  &   0.028  &   0.004  &   75  &        &         \\  
NGC 237          &  35   &    33    &   49  &    179   &    8.62  &   -0.302  &   0.037  &   0.024  &  171  &  0.2   &   1887  \\  
                 &  32.9 &    33.1  &   45  &      1   &    8.62  &   -0.302  &   0.034  &   0.003  &   48  &        &         \\  
NGC 444          &  36   &    31    &   75  &    160   &    8.49  &   -0.207  &   0.050  &   0.019  &   51  &        &    813  \\  
                 &  36.9 &    31.9  &   71  &    158   &    8.49  &   -0.240  &   0.050  &   0.006  &   48  &        &         \\  
NGC 477          &  33   &    30    &   57  &    142   &    8.63  &   -0.236  &   0.045  &   0.023  &    9  &        &    939  \\  
                 &  35.3 &    31.3  &   56  &    141   &    8.63  &   -0.249  &   0.044  &   0.011  &   39  &        &         \\  
NGC 496          &  35   &    32    &   54  &     35   &    8.60  &   -0.276  &   0.047  &   0.054  &   18  &        &   1033  \\  
                 &  38.1 &    33.1  &   58  &     33   &    8.62  &   -0.336  &   0.039  &   0.004  &  135  &        &         \\  
NGC 776          &  39   &    34    &   21  &     55   &    8.62  &   -0.075  &   0.035  &   0.022  &   30  &        &   1032  \\  
                 &  32.7 &    29.7  &   29  &     19   &    8.62  &   -0.073  &   0.033  &   0.007  &   99  &        &         \\  
NGC 1542         &  37   &    33    &   66  &    129   &    8.56  &   -0.114  &   0.029  &   0.017  &   78  &        &    213  \\  
                 &  38.9 &    35.9  &   72  &    126   &    8.57  &   -0.151  &   0.027  &   0.009  &   27  &        &         \\  
NGC 2253         &  40   &    33    &   43  &    127   &    8.62  &   -0.120  &   0.027  &   0.022  &   81  &        &   1275  \\  
                 &  41.3 &    38.3  &   42  &    140   &    8.62  &   -0.133  &   0.024  &   0.003  &  150  &        &         \\  
NGC 2347         &  39   &    32    &   52  &      4   &    8.67  &   -0.329  &   0.038  &   0.030  &  162  &  0.3   &   1306  \\  
                 &  36.7 &    32.9  &   51  &      4   &    8.67  &   -0.338  &   0.034  &   0.005  &  135  &        &         \\  
NGC 2410         &  37   &    32    &   72  &     35   &    8.62  &   -0.119  &   0.025  &   0.016  &  123  &        &    545  \\  
                 &  40.5 &    26.7  &   75  &     34   &    8.62  &   -0.105  &   0.024  &   0.004  &  120  &        &         \\  
NGC 2730         &  38   &    31    &   43  &     86   &    8.51  &   -0.146  &   0.037  &   0.008  &    3  &        &   2689  \\  
                 &  36.5 &    30.9  &   46  &     86   &    8.51  &   -0.144  &   0.037  &   0.004  &   75  &        &         \\  
NGC 2906         &  38   &    33    &   55  &     80   &    8.61  &   -0.052  &   0.021  &   0.005  &   99  &        &   1574  \\  
                 &  39.1 &    33.9  &   67  &     73   &    8.61  &   -0.044  &   0.020  &   0.003  &  165  &        &         \\  
NGC 2916         &  37   &    33    &   51  &     17   &    8.63  &   -0.236  &   0.032  &   0.009  &  153  &        &   1028  \\  
                 &  36.5 &    32.1  &   45  &     12   &    8.64  &   -0.267  &   0.031  &   0.006  &  156  &        &         \\  
NGC 3381         &  38   &    32    &   28  &     80   &    8.52  &   -0.191  &   0.037  &   0.025  &  150  &        &   2914  \\  
                 &  33.3 &    33.9  &   32  &     94   &    8.53  &   -0.193  &   0.035  &   0.003  &  147  &        &         \\  
NGC 3614         &  38   &    33    &   55  &    101   &    8.59  &   -0.249  &   0.030  &   0.013  &  156  &        &   1218  \\  
                 &  37.1 &    34.3  &   56  &    100   &    8.59  &   -0.254  &   0.029  &   0.005  &   27  &        &         \\  
NGC 3811         &  38   &    33    &   35  &    169   &    8.63  &   -0.196  &   0.024  &   0.007  &   81  &        &   1909  \\  
                 &  38.3 &    31.7  &   37  &    182   &    8.63  &   -0.192  &   0.024  &   0.004  &   84  &        &         \\  
NGC 3994         &  37   &    33    &   59  &      7   &    8.55  &   -0.059  &   0.027  &   0.009  &  120  &        &   1496  \\  
                 &  36.5 &    34.1  &   27  &    132   &    8.55  &   -0.078  &   0.026  &   0.007  &   60  &        &         \\  
NGC 4185         &  38   &    33    &   49  &    166   &    8.61  &   -0.135  &   0.036  &   0.020  &  144  &        &    416  \\  
                 &  34.9 &    33.1  &   66  &    161   &    8.61  &   -0.085  &   0.034  &   0.009  &   21  &        &         \\  
NGC 4210         &  39   &    32    &   41  &     96   &    8.64  &   -0.192  &   0.024  &   0.010  &   36  &        &   1009  \\  
                 &  38.1 &    31.5  &   40  &     88   &    8.63  &   -0.188  &   0.024  &   0.005  &  114  &        &         \\  
NGC 4470         &  34   &    36    &   48  &      0   &    8.47  &   -0.125  &   0.036  &   0.014  &   42  &        &   1680  \\  
                 &  31.5 &    35.9  &   26  &     20   &    8.48  &   -0.167  &   0.035  &   0.005  &    0  &        &         \\  
NGC 4644         &  39   &    34    &   71  &     52   &    8.62  &   -0.084  &   0.029  &   0.011  &  174  &        &    597  \\  
                 &  37.3 &    35.3  &   64  &     40   &    8.63  &   -0.113  &   0.029  &   0.003  &   57  &        &         \\  
NGC 5205         &  39   &    33    &   54  &    165   &    8.55  &   -0.020  &   0.036  &   0.010  &   93  &        &    381  \\  
                 &  30.5 &    31.3  &   58  &    138   &    8.56  &   -0.030  &   0.035  &   0.007  &   84  &        &         \\  
NGC 5406         &  37   &    33    &   43  &    122   &    8.64  &   -0.147  &   0.035  &   0.010  &  141  &        &   1050  \\  
                 &  38.9 &    33.5  &   52  &    109   &    8.64  &   -0.130  &   0.034  &   0.006  &   60  &        &         \\  
NGC 5520         &  39   &    33    &   59  &     65   &    8.60  &   -0.189  &   0.032  &   0.011  &  171  &        &   2316  \\  
                 &  37.7 &    32.9  &   50  &     70   &    8.60  &   -0.218  &   0.030  &   0.002  &  129  &        &         \\  
NGC 5614         &  39   &    33    &   34  &    134   &    8.64  &   -0.184  &   0.041  &   0.014  &   63  &        &    741  \\  
                 &  41.1 &    35.3  &   41  &    138   &    8.64  &   -0.169  &   0.040  &   0.006  &  150  &        &         \\  
NGC 5720         &  39   &    34    &   47  &    131   &    8.62  &   -0.186  &   0.048  &   0.018  &  168  &  0.3   &    344  \\  
                 &  36.7 &    33.1  &   45  &    112   &    8.63  &   -0.204  &   0.046  &   0.009  &    0  &        &         \\  
   \hline
\end{tabular}\\
\end{center}
\end{table*}

\setcounter{table}{1}
\begin{table*}
\caption[]{
Continued
}
\begin{center}
\begin{tabular}{lllcrccccccr} \hline \hline
Name                                          &
\multicolumn{2}{c}{Center location}           &  
$i$                                          &
P.A.                                          &
(O/H)$_{0}^{a}$                                & 
O/H gradient                                  &
$\sigma$(O/H)                                 &
\multicolumn{2}{c}{Maximum azimuthal asymmetry}  &  
Break                                        &
Number of                                    \\
                                        &
$X_{0}$                                  &
$Y_{0}$                                  &
                                        &
                                        &
                                        & 
                                        &
                                        &
amplitude                               &
angle                                   &
radius                                  &
spectra                                 \\

                             &
spaxel                        &
spaxel                        &
degree                       &
degree                       &
                             &
dex $R_{25}^{-1}$              &
dex                          & 
dex                          &
degree                       &
$R_{25}$                      &
                             \\  \hline 
1                            &
2                            &
3                            &
4                            &
5                            &
6                            &
7                            &
8                            & 
9                            &
10                           &
11                           &
12                           \\  \hline   
NGC 5888         &  38   &    32    &   52  &    154   &    8.61  &   -0.098  &   0.044  &   0.040  &   69  &        &    299  \\  
                 &  45.5 &    44.5  &   67  &    149   &    8.61  &   -0.067  &   0.034  &   0.010  &  174  &        &         \\  
NGC 5947         &  43   &    36    &   31  &     63   &    8.66  &   -0.272  &   0.035  &   0.015  &   93  &        &   1404  \\  
                 &  42.5 &    34.7  &   30  &     55   &    8.66  &   -0.267  &   0.034  &   0.005  &   45  &        &         \\  
NGC 6004         &  38   &    33    &   34  &     87   &    8.63  &   -0.161  &   0.036  &   0.012  &   66  &        &   1578  \\  
                 &  36.1 &    31.5  &   43  &     90   &    8.63  &   -0.146  &   0.036  &   0.006  &    3  &        &         \\  
NGC 6063         &  38   &    34    &   56  &    156   &    8.58  &   -0.197  &   0.043  &   0.026  &  150  &        &   1795  \\  
                 &  36.3 &    35.7  &   58  &    149   &    8.57  &   -0.176  &   0.037  &   0.007  &  126  &        &         \\  
NGC 6132         &  36   &    34    &   68  &    125   &    8.61  &   -0.306  &   0.051  &   0.046  &  147  &  0.2   &    875  \\  
                 &  34.3 &    34.3  &   69  &    126   &    8.63  &   -0.332  &   0.044  &   0.008  &   93  &        &         \\  
NGC 6154         &  39   &    35    &   31  &    153   &    8.63  &   -0.152  &   0.026  &   0.010  &  159  &        &    438  \\  
                 &  37.1 &    37.3  &   37  &    143   &    8.64  &   -0.175  &   0.025  &   0.005  &   90  &        &         \\  
NGC 6478         &  38   &    34    &   68  &     34   &    8.60  &   -0.157  &   0.034  &   0.014  &   66  &        &   1120  \\  
                 &  37.3 &    33.5  &   71  &     37   &    8.59  &   -0.134  &   0.034  &   0.005  &  114  &        &         \\  
NGC 6978         &  38   &    33    &   67  &    128   &    8.60  &   -0.077  &   0.034  &   0.033  &   48  &        &    176  \\  
                 &  41.3 &    38.5  &   62  &    126   &    8.62  &   -0.141  &   0.029  &   0.007  &   51  &        &         \\  
NGC 7311         &  38   &    33    &   60  &     12   &    8.62  &   -0.128  &   0.029  &   0.011  &  108  &        &    990  \\  
                 &  38.1 &    35.7  &   66  &     11   &    8.62  &   -0.126  &   0.028  &   0.001  &   18  &        &         \\  
NGC 7321         &  37   &    33    &   46  &     13   &    8.62  &   -0.200  &   0.033  &   0.008  &   81  &  0.3   &   1745  \\  
                 &  36.9 &    32.3  &   48  &     12   &    8.62  &   -0.197  &   0.032  &   0.003  &  171  &        &         \\  
NGC 7466         &  39   &    32    &   66  &     25   &    8.57  &   -0.109  &   0.032  &   0.023  &  123  &        &    798  \\  
                 &  37.7 &    36.7  &   59  &     28   &    8.56  &   -0.110  &   0.029  &   0.003  &   96  &        &         \\  
NGC 7489         &  38   &    32    &   54  &    160   &    8.69  &   -0.628  &   0.062  &   0.034  &  144  &  0.2   &   1843  \\  
                 &  37.7 &    33.3  &   59  &    162   &    8.69  &   -0.592  &   0.058  &   0.008  &  117  &        &         \\  
NGC 7549         &  38   &    34    &   67  &     18   &    8.58  &   -0.092  &   0.030  &   0.006  &  141  &        &    895  \\  
                 &  36.7 &    34.5  &   23  &    164   &    8.59  &   -0.177  &   0.027  &   0.004  &  174  &        &         \\  
NGC 7591         &  38   &    33    &   54  &    156   &    8.63  &   -0.165  &   0.031  &   0.014  &   12  &        &    991  \\  
                 &  36.7 &    33.7  &   49  &    151   &    8.63  &   -0.194  &   0.031  &   0.004  &   27  &        &         \\  
NGC 7625         &  36   &    32    &   30  &     22   &    8.59  &   -0.091  &   0.028  &   0.010  &  156  &        &   1620  \\  
                 &  33.7 &    32.7  &   45  &    136   &    8.60  &   -0.101  &   0.027  &   0.003  &   96  &        &         \\  
NGC 7631         &  39   &    33    &   65  &     76   &    8.64  &   -0.198  &   0.033  &   0.009  &   12  &  0.2   &   1017  \\  
                 &  40.7 &    32.7  &   67  &     75   &    8.64  &   -0.184  &   0.033  &   0.004  &   18  &        &         \\  
NGC 7653         &  37   &    33    &   31  &    175   &    8.63  &   -0.228  &   0.033  &   0.010  &  120  &        &   2226  \\  
                 &  35.9 &    34.1  &   21  &    164   &    8.63  &   -0.242  &   0.032  &   0.004  &  162  &        &         \\  
NGC 7716         &  37   &    36    &   34  &     41   &    8.52  &   -0.058  &   0.032  &   0.005  &   78  &        &   1965  \\  
                 &  35.3 &    36.9  &   61  &     15   &    8.52  &   -0.032  &   0.032  &   0.004  &   87  &        &         \\  
NGC 7819         &  34   &    34    &   55  &    104   &    8.60  &   -0.267  &   0.050  &   0.031  &   99  &        &   1184  \\  
                 &  32.3 &    32.1  &   56  &     92   &    8.60  &   -0.269  &   0.044  &   0.008  &  177  &        &         \\  
IC  674          &  38   &    32    &   66  &    121   &    8.59  &   -0.157  &   0.045  &   0.028  &  141  &  0.2   &    386  \\  
                 &  37.7 &    33.7  &   70  &    120   &    8.58  &   -0.120  &   0.041  &   0.013  &    3  &        &         \\  
IC  1199         &  38   &    33    &   68  &    164   &    8.62  &   -0.139  &   0.029  &   0.015  &   78  &        &    771  \\  
                 &  38.5 &    35.5  &   63  &    162   &    8.62  &   -0.164  &   0.028  &   0.003  &   33  &        &         \\  
IC  1256         &  38   &    35    &   52  &     92   &    8.66  &   -0.273  &   0.040  &   0.011  &  141  &        &   1219  \\  
                 &  37.3 &    34.9  &   51  &     88   &    8.66  &   -0.277  &   0.040  &   0.008  &   18  &        &         \\  
IC  1528         &  39   &    33    &   67  &     74   &    8.62  &   -0.319  &   0.048  &   0.017  &    3  &        &   1934  \\  
                 &  37.5 &    33.1  &   66  &     76   &    8.62  &   -0.331  &   0.047  &   0.006  &  162  &        &         \\  
IC  2487         &  37   &    35    &   75  &    164   &    8.61  &   -0.235  &   0.038  &   0.023  &  150  &        &    867  \\  
                 &  36.5 &    36.3  &   76  &    164   &    8.61  &   -0.236  &   0.034  &   0.007  &   60  &        &         \\  
UGC 5            &  38   &    32    &   60  &     46   &    8.59  &   -0.133  &   0.032  &   0.012  &  126  &  0.2   &   1189  \\  
                 &  36.5 &    33.9  &   60  &     46   &    8.59  &   -0.128  &   0.031  &   0.005  &  177  &        &         \\  
UGC 809          &  32   &    31    &   77  &     24   &    8.46  &   -0.123  &   0.047  &   0.038  &  171  &  0.3   &    633  \\  
                 &  34.1 &    30.5  &   72  &     28   &    8.48  &   -0.188  &   0.040  &   0.005  &   75  &        &         \\  
UGC 1057         &  36   &    32    &   71  &    152   &    8.63  &   -0.312  &   0.047  &   0.040  &  126  &  0.2   &    957  \\  
                 &  35.9 &    33.7  &   68  &    156   &    8.63  &   -0.325  &   0.042  &   0.008  &   51  &        &         \\  
UGC 1938         &  35   &    29    &   73  &    155   &    8.62  &   -0.284  &   0.045  &   0.043  &   60  &  0.2   &    679  \\  
                 &  36.9 &    31.9  &   71  &    155   &    8.64  &   -0.338  &   0.040  &   0.008  &   30  &        &         \\  
UGC 2403         &  33   &    32    &   67  &    152   &    8.65  &   -0.181  &   0.035  &   0.029  &  153  &        &    631  \\  
                 &  31.1 &    33.1  &   57  &    151   &    8.66  &   -0.224  &   0.029  &   0.005  &   69  &        &         \\  
UGC 3107         &  36   &    32    &   76  &     73   &    8.60  &   -0.141  &   0.030  &   0.016  &  162  &        &    251  \\  
                 &  34.9 &    32.7  &   74  &     67   &    8.59  &   -0.138  &   0.026  &   0.007  &  129  &        &         \\  
   \hline
\end{tabular}\\
\end{center}
\end{table*}

\setcounter{table}{1}
\begin{table*}
\caption[]{
Continued
}
\begin{center}
\begin{tabular}{lllcrccccccr} \hline \hline
Name                                          &
\multicolumn{2}{c}{Center location}           &  
$i$                                          &
P.A.                                          &
(O/H)$_{0}^{a}$                                & 
O/H gradient                                  &
$\sigma$(O/H)                                 &
\multicolumn{2}{c}{Maximum azimuthal asymmetry}  &  
Break                                        &
Number of                                    \\
                                        &
$X_{0}$                                  &
$Y_{0}$                                  &
                                        &
                                        &
                                        & 
                                        &
                                        &
amplitude                               &
angle                                   &
radius                                  &
spectra                                 \\

                             &
spaxel                        &
spaxel                        &
degree                       &
degree                       &
                             &
dex $R_{25}^{-1}$              &
dex                          & 
dex                          &
degree                       &
$R_{25}$                      &
                             \\  \hline 
1                            &
2                            &
3                            &
4                            &
5                            &
6                            &
7                            &
8                            & 
9                            &
10                           &
11                           &
12                           \\  \hline   
UGC 3253         &  42   &    36    &   55  &     81   &    8.72  &   -0.354  &   0.031  &   0.013  &   57  &  0.3   &    622  \\  
                 &  41.7 &    35.9  &   56  &     79   &    8.72  &   -0.352  &   0.030  &   0.009  &   54  &        &         \\  
UGC 3969         &  36   &    32    &   78  &    135   &    8.57  &   -0.091  &   0.025  &   0.006  &   18  &        &    320  \\  
                 &  36.5 &    30.9  &   56  &    135   &    8.57  &   -0.117  &   0.021  &   0.004  &  129  &        &         \\  
UGC 4132         &  37   &    32    &   73  &     28   &    8.63  &   -0.147  &   0.021  &   0.015  &   72  &        &   1041  \\  
                 &  37.1 &    33.7  &   70  &     29   &    8.63  &   -0.169  &   0.019  &   0.004  &  153  &        &         \\  
UGC 5396         &  38   &    31    &   70  &    156   &    8.61  &   -0.261  &   0.040  &   0.034  &   45  &        &    338  \\  
                 &  36.3 &    30.5  &   66  &    151   &    8.63  &   -0.320  &   0.036  &   0.010  &    6  &        &         \\  
UGC 5598         &  38   &    32    &   73  &     36   &    8.56  &   -0.148  &   0.033  &   0.014  &   18  &        &    621  \\  
                 &  37.1 &    32.3  &   69  &     35   &    8.56  &   -0.167  &   0.032  &   0.006  &  156  &        &         \\  
UGC 8107         &  39   &    33    &   68  &     52   &    8.57  &   -0.203  &   0.046  &   0.048  &  144  &        &    851  \\  
                 &  35.7 &    35.3  &   63  &     40   &    8.58  &   -0.232  &   0.037  &   0.006  &  126  &        &         \\  
UGC 8778         &  36   &    32    &   74  &    118   &    8.55  &   -0.067  &   0.025  &   0.019  &  135  &        &    289  \\  
                 &  37.5 &    34.9  &   79  &    124   &    8.56  &   -0.070  &   0.022  &   0.004  &   39  &        &         \\  
UGC 9067         &  39   &    32    &   62  &     13   &    8.66  &   -0.346  &   0.045  &   0.033  &  162  &  0.2   &    980  \\  
                 &  38.3 &    32.9  &   56  &     11   &    8.68  &   -0.400  &   0.041  &   0.007  &   54  &        &         \\  
UGC 9476         &  37   &    33    &   48  &    125   &    8.55  &   -0.126  &   0.023  &   0.008  &  120  &        &   1220  \\  
                 &  37.9 &    33.3  &   53  &    134   &    8.55  &   -0.118  &   0.022  &   0.004  &  114  &        &         \\  
UGC 9665         &  38   &    33    &   76  &    143   &    8.50  &   -0.086  &   0.038  &   0.018  &  159  &        &    970  \\  
                 &  37.5 &    32.3  &   62  &    112   &    8.50  &   -0.092  &   0.031  &   0.008  &   54  &        &         \\  
UGC 9873         &  38   &    35    &   76  &    125   &    8.55  &   -0.140  &   0.034  &   0.022  &  177  &        &    534  \\  
                 &  37.3 &    35.7  &   73  &    125   &    8.55  &   -0.159  &   0.032  &   0.005  &   36  &        &         \\  
UGC 9892         &  38   &    34    &   73  &    101   &    8.59  &   -0.177  &   0.028  &   0.012  &  105  &        &    616  \\  
                 &  37.5 &    33.7  &   72  &    101   &    8.59  &   -0.182  &   0.027  &   0.004  &   54  &        &         \\  
UGC 10331        &  38   &    34    &   77  &    142   &    8.46  &   -0.167  &   0.054  &   0.046  &   45  &        &   1063  \\  
                 &  35.7 &    32.5  &   61  &    142   &    8.47  &   -0.223  &   0.047  &   0.008  &  165  &        &         \\  
UGC 10384        &  39   &    33    &   76  &     91   &    8.59  &   -0.170  &   0.042  &   0.018  &   87  &        &    924  \\  
                 &  39.1 &    34.3  &   57  &     94   &    8.60  &   -0.244  &   0.041  &   0.018  &    3  &        &         \\  
UGC 10710        &  39   &    36    &   75  &    148   &    8.58  &   -0.134  &   0.030  &   0.007  &  156  &        &    835  \\  
                 &  38.9 &    36.3  &   77  &    148   &    8.59  &   -0.133  &   0.030  &   0.003  &  159  &        &         \\  
UGC 10811        &  39   &    33    &   68  &     93   &    8.66  &   -0.250  &   0.041  &   0.020  &  144  &  0.4   &    337  \\  
                 &  36.7 &    33.1  &   70  &     88   &    8.68  &   -0.259  &   0.040  &   0.012  &   87  &        &         \\  
UGC 10972        &  37   &    32    &   75  &     54   &    8.62  &   -0.237  &   0.039  &   0.016  &    6  &  0.2   &    806  \\  
                 &  35.9 &    32.7  &   77  &     53   &    8.62  &   -0.218  &   0.036  &   0.008  &  123  &        &         \\  
UGC 11262        &  37   &    33    &   68  &     48   &    8.57  &   -0.269  &   0.064  &   0.040  &  114  &        &    568  \\  
                 &  36.3 &    34.7  &   60  &     53   &    8.58  &   -0.304  &   0.060  &   0.011  &   60  &        &         \\  
UGC 11649        &  38   &    33    &   38  &     70   &    8.65  &   -0.231  &   0.040  &   0.016  &   90  &  0.2   &    424  \\  
                 &  38.5 &    31.1  &   33  &     58   &    8.65  &   -0.225  &   0.040  &   0.005  &   93  &        &         \\  
UGC 11792        &  40   &    32    &   77  &    160   &    8.56  &   -0.105  &   0.030  &   0.019  &  177  &        &    420  \\  
                 &  39.3 &    32.1  &   79  &    160   &    8.56  &   -0.109  &   0.028  &   0.005  &   60  &        &         \\  
UGC 11982        &  39   &    31    &   78  &    169   &    8.44  &   -0.122  &   0.062  &   0.068  &  153  &        &    309  \\  
                 &  37.7 &    33.5  &   74  &    164   &    8.48  &   -0.205  &   0.053  &   0.024  &   48  &        &         \\  
UGC 12185        &  39   &    33    &   62  &    147   &    8.56  &   -0.139  &   0.044  &   0.011  &   39  &        &    498  \\  
                 &  37.7 &    33.3  &   68  &    157   &    8.57  &   -0.146  &   0.042  &   0.009  &  105  &        &         \\  
UGC 12224        &  38   &    32    &   24  &     47   &    8.59  &   -0.277  &   0.038  &   0.009  &   96  &        &   1069  \\  
                 &  37.9 &    32.5  &   28  &     28   &    8.59  &   -0.274  &   0.038  &   0.007  &   96  &        &         \\  
UGC 12519        &  38   &    33    &   75  &    158   &    8.54  &   -0.129  &   0.030  &   0.024  &  114  &        &   1217  \\  
                 &  37.7 &    35.7  &   71  &    155   &    8.55  &   -0.161  &   0.028  &   0.005  &   69  &        &         \\  
UGC 12816        &  39   &    34    &   53  &    140   &    8.51  &   -0.336  &   0.062  &   0.034  &    9  &        &    716  \\  
                 &  36.9 &    32.9  &   56  &    144   &    8.52  &   -0.328  &   0.059  &   0.008  &  102  &        &         \\  
UGC 12864        &  37   &    35    &   64  &    106   &    8.51  &   -0.219  &   0.055  &   0.025  &   60  &        &    953  \\  
                 &  35.9 &    33.7  &   63  &     87   &    8.51  &   -0.240  &   0.048  &   0.010  &  105  &        &         \\  
PGC 1841         &  34   &    32    &   66  &    165   &    8.59  &   -0.092  &   0.030  &   0.013  &  138  &        &    989  \\  
                 &  33.1 &    33.9  &   70  &    169   &    8.60  &   -0.091  &   0.029  &   0.003  &   93  &        &         \\  
PGC 64373        &  39   &    32    &   67  &    159   &    8.65  &   -0.242  &   0.036  &   0.035  &  159  &        &   1060  \\  
                 &  37.9 &    33.5  &   67  &    159   &    8.67  &   -0.282  &   0.031  &   0.005  &   75  &        &         \\  
   \hline
\end{tabular}\\
\end{center}
\begin{flushleft}
$^a$  in units of 12 + log(O/H)   \\
\end{flushleft}
\end{table*}

\end{document}